\documentclass[iop]{emulateapj}
\usepackage{soul}
\usepackage{lineno}
\usepackage{aas_macros}

\usepackage[colorlinks,citecolor=blue]{hyperref}%backref,

\usepackage{amssymb}

\usepackage{enumitem}
\usepackage{mathtools}
\usepackage[cm]{sfmath}

\usepackage{multirow}

\usepackage{listings}
\usepackage{csvsimple}
\usepackage{longtable}
\usepackage{array}
\usepackage{chapterbib}
\usepackage{subfiles}
\begin{document}

%Entropy and Melting implications

\title{Silicate Melting and Vaporization\\during Rocky Planet Formation}

\author{E. J. Davies$^{1}$, P. J. Carter$^{1}$, S. Root$^{2}$, R. G. Kraus$^{3}$, D. K. Spaulding$^{1}$, S.~T.~Stewart$^{1}$, S. B. Jacobsen$^{4}$ }

\affil{$^{1}$Department of Earth and Planetary Sciences, University of California Davis, One Shields Avenue, Davis, CA 95616, USA\\
   $^{2}$Sandia National Laboratories, Albuquerque, NM, USA\\
   $^{3}$Lawrence Livermore National Laboratory, Livermore, CA, USA\\
   $^{4}$Department of Earth and Planetary Sciences, Harvard University, Cambridge, MA 02138, USA\vspace{1mm}\\
   {\it Accepted for publication in JGR: Planets}}
\email{ejdavies@ucdavis.edu}

%% 3 Keypoints, final entry on title page.
%% each one limited to 140 characters

%% \begin{abstract} starts the second page 

\begin{abstract}
Collisions that induce melting and vaporization can have a substantial effect on the thermal and geochemical evolution of planets. However, the thermodynamics of major minerals are not well known at the extreme conditions attained during planet formation. We obtained new data at the Sandia Z Machine and use published thermodynamic data for the major mineral forsterite (Mg$_2$SiO$_4$) to calculate the specific entropy in the liquid region of the principal Hugoniot. We use our calculated specific entropy of shocked forsterite, and revised entropies for shocked silica, to determine the critical impact velocities for melting or vaporization upon decompression from the shocked state to 1 bar and the triple points, which are near the pressures of the solar nebula. We also demonstrate the importance of the initial temperature on the criteria for vaporization. 
Applying these results to $N$-body simulations of terrestrial planet formation, we find that {up to 20 to 40\%} of the total system mass is processed through collisions with velocities that exceed the criteria for incipient vaporization at the triple point. Vaporizing collisions between small bodies {are} an important component of terrestrial planet formation.\\

{\small KEY POINTS:}
\begin{itemize}[itemsep=0mm,leftmargin=17mm,rightmargin=15mm]
\vspace{-2mm}
\item We calculate specific entropy on the principal Hugoniot of forsterite (Mg$_2$SiO$_4$) by thermodynamic integration of experimental data.
\item We derive new criteria for shock-induced melting and vaporization.
\item Vaporizing collisions between small bodies are an important component of terrestrial planet formation.\\
\vspace{-3mm}

\end{itemize}

% 198/200 word limit
{\small PLAIN LANGUAGE SUMMARY:}\\
During planet formation, collisions onto planets and between planetary building blocks, such as asteroids, can be fast enough to melt or vaporize rock. Melting and vaporization changes the chemical make-up of planets. However, until recently, the extreme pressures and temperatures reached during planetary collisions could not be reproduced in laboratory experiments. We were missing key measurements on major materials that make up Earth's mantle, such as the mineral forsterite (Mg$_2$SiO$_4$). Here, we used the Z Machine, a facility at Sandia National Laboratories that can launch projectiles up to 40 km~s$^{-1}$ (almost 90,000 miles per hour), to measure the properties of forsterite at extreme conditions. Based on these measurements, we calculated that collisions faster than {8}.2 km~s$^{-1}$ (about 1{8},000 miles per hour) can completely melt and begin to vaporize the rocky portions of planets and their building blocks. We then analyzed computer simulations {of planet formation} to determine how much material could have been melted or vaporized during the growth of our rocky planets. We found that {20 to 40\%} of all the material that makes up the inner solar system could have been involved in collisions that melted and vaporized rock.\\

%(187/250 words for JGR)
\end{abstract}


%% ------------------------------------------------------------------------ %%
%
%  TEXT
%
%% ------------------------------------------------------------------------ %%https://www.overleaf.com/project/5b1824ea35b6702b8d817b91

\section{Introduction}
Collisions are a key aspect of planet formation \citep[e.g.][]{chambers2010terrestrial}. The outcomes of collisions are diverse and complex, ranging from growth of planetesimals to the formation of synestias
\citep{leinhardt2012, lock2017structure}.
Because collisions deposit energy and redistribute material, they can significantly affect the thermal and geochemical evolution of growing planets \citep{stewart2012collisions, asphaug2010similar,carter2015compositional,carter2018collisional}. 
The increase in internal energy from a collision is generated by the shock from the initial impact and secondary shocks because of the changes in the gravitational potential well.
The impact conditions required to reach the onset of melting and vaporization depend on the equations of state (EOS) and ambient conditions of the constituent materials \citep{stewart2005shock, kraus2012shock, kraus2015impact}.

Forsterite (Mg$_2$SiO$_4$), the magnesium end-member of the olivine system, is a major silicate phase among the first solids in the solar nebula \citep{lodders2003solar}, and olivine ((Mg,Fe)$_2$SiO$_4$) is an abundant phase in primitive meteorites. Thus, olivine is a major phase in the mantles of differentiated planetesimals and planets. Because terrestrial olivines are rich in Mg, with compositions near ((Mg$_{0.9}$Fe$_{0.1}$)$_2$SiO$_4$), and there is an abundance of data on forsterite, simulations of planetary collisions often use forsterite as a proxy for the bulk silicate composition of differentiated bodies. {While bridgmanite, (Mg,Fe)SiO$_3$, is the primary silicate phase in the lower mantle of the Earth, the upper mantle and planetary building blocks are likely to be dominated by olivines. }Hence, the criteria for melting and vaporization of forsterite form important constraints for the amount of impact-induced melting and vaporization during planet formation. Although an analysis based on single component systems neglects the complexity of incongruent melting and vaporization in multi-component systems, at this time, impact-induced phase changes in single component systems can be calculated more robustly than in multi-component systems.

Currently, one of the most widely used equations of state for forsterite among impact modelers is an extension of the ANEOS (Analytic Equations of State) model \citep{thompson1974improvements,thompson1990aneos} to include molecular vapor species (M-ANEOS), which was developed for silica in \citet{melosh2007hydrocode}. 
ANEOS is a collection of analytic expressions that describe the Helmholtz free energy across a wide range of pressures and temperatures. ANEOS requires approximately 40 input parameters to describe a material, and there are a few different sets of input parameters for forsterite in current use for impact modeling \citep{canup2012forming,canup2013lunar,collins2014improvements,nakajima2014investigation,cuk2012making}. These parameter sets were developed before experimental data above 250 GPa was available. Recent measurements of the principal shock Hugoniot of forsterite \citep{root2018forsterite} found that the previous ANEOS models diverge from the data in the liquid region.  \citet{stewart2019improvements} have modified ANEOS to improve the fit in the liquid region and provide a revised set of input parameters for forsterite. Although the free energy expressions in ANEOS are sometimes poor approximations of the underlying physics, it has the capability to generate thermodynamically consistent EOS models over the extremely wide range of pressures and temperatures that are vital for numerical simulations of planetary collisions.

Our goal is to describe the thermodynamics of liquid forsterite. Recently, the principal Hugoniot of forsterite has been measured at the Z machine at Sandia and the Omega laser at the U. Rochester up to 950 GPa \citep{root2018forsterite}. These measurements provide pressure, specific volume, and temperature (P-V-T) along a single line on the EOS surface. However, to develop a wide-ranging EOS, we need other thermodynamic information. One of the most important variables to predict phase changes is specific entropy ($S$). Here, we present an experimentally constrained thermodynamic integration to calculate specific entropy on the principal Hugoniot. The Gr\"uneisen parameter is needed to determine thermodynamic states off the principal Hugoniot. We present new measurements of the Gr\"uneisen parameter obtained from shallow release experiments on the Sandia Z machine.

\begin{figure*}
\centering
\includegraphics[width=5in]{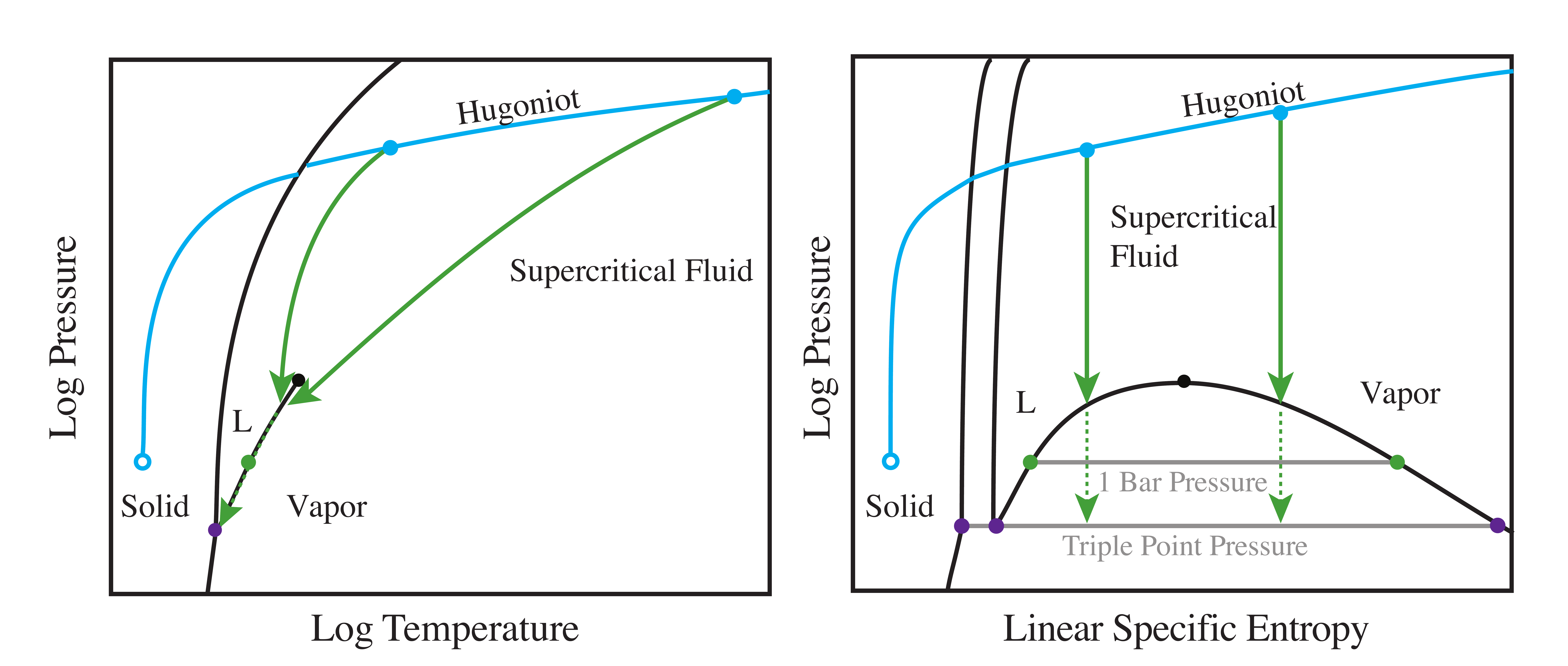}

\caption{Schematic of a generalized single component phase diagram in $T$-$P$ and $S$-$P$ space. The solid, liquid (L), and vapor phase boundaries (black lines) are shown, along with the critical point (black point), triple point (purple points), and 1 bar pressure (green points). The blue curve is the Hugoniot, the locus of possible states{ reached by single shocks into material in a given initial state at 1 bar pressure and a particular temperature}. The green lines show decompression along isentropes from specific shock states (blue points). (Left) In temperature-pressure space, when the decompression isentropes intercept the liquid-vapor phase boundary, the bulk material is a mixture of liquid and vapor, however it is difficult to determine the mass fraction of each phase. (Right) In $S$-$P$ space, the liquid-vapor phase boundary is represented by a dome that shows the specific entropy difference between liquid and vapor in the mixed phase region. The triple point pressure of silicates is similar to the fiducial pressure of the solar nebula at 1 au, about 10$^{-4}$ bar.}
\label{fig:ent_schem}
\end{figure*}

To apply our work to planetary collisions, we use the {\em entropy method} to calculate phase changes after shock compression and isentropic decompression \citep{ahrens1972shock}. Figure \ref{fig:ent_schem} presents a schematic of the entropy method and the thermodynamic path of a parcel of shocked material. The shock wave increases the pressure, temperature, and specific entropy to a state on the Hugoniot. The shocked material will decompress from the Hugoniot to the pressure of the surrounding medium via a rarefaction wave. This rarefaction wave propagates at the speed of sound, is faster than thermal diffusion, and does no work on the material. Consequently, {decompression via a rarefaction wave is approximately isentropic, and the isentropic assumption produces an absolute lower bound to the entropy of the release state. When there is no shear strength, as in a liquid, the decompression path is reversible and isentropic}. Thus, the specific entropy of the shocked state corresponds to the specific entropy of the decompressed state. Some estimates of shock-induced melting and vaporization have reported the criteria for phase changes on decompression into Earth's atmosphere at 10$^5$ Pa (1 bar). However, during accretion, the pressure of the protoplanetary nebula is more appropriate, e.g., 1 to 10 Pa at 1 au \citep{wood2000pressure}. These pressures are close to the triple points of silicates, approximately 2 Pa for silica \citep{mysen1988condensation} and 5.2 Pa for forsterite \citep{nagahara1994evaporation}. When the final specific entropy falls in a mixed phase region, the lever rule is used to determine the mass fraction of each phase of material. The lever rule defines the mass fraction of the second phase as
\begin{linenomath*}
\begin{equation}
    X_2=\frac{S_B-S_1}{S_2-S_1},
    \label{eq:lever}
\end{equation}
\end{linenomath*}
where $S_B$ is the specific entropy of the bulk parcel, $S_1$ is the specific entropy of the first phase, and $S_2$ is the specific entropy of the second phase. In the case of a decompression into a liquid-vapor mixture, the first phase is liquid, the second phase is vapor, and equation \ref{eq:lever} gives the mass fraction of the parcel that is in the vapor phase.

Previous studies estimated shock-induced melting and vaporization of various silicates using the entropy method: (Mg$_{0.88}$Fe$_{0.12}$)$_2$SiO$_4$ \citep{tonks1993magma}, Mg$_2$SiO$_4$ \citep{pierazzo1997reevaluation}, and SiO$_2$ \citep{kraus2012shock}. In the case of \citet{tonks1993magma} and \citet{pierazzo1997reevaluation}, melting and vaporization was estimated using an equation of state model constructed with limited data for the vapor curve. For silica, \citet{kraus2012shock} determined specific entropy on the principal Hugoniot via thermodynamic integration and constrained the liquid-vapor phase boundary from experimental data at 1 bar. Our work applies the techniques developed in \citet{kraus2012shock} to derive wide-ranging EOS information.

{We also investigate the process of impact-induced melting and vaporization in the context of terrestrial planet accretion. In general, numerical simulations of planet formation have focused on capturing the mechanical outcomes of collisions while neglecting the details of the thermal effects of collisions.} $N$-body {techniques} are used to investigate the accretion of planets, and many studies assumed that each collision results in perfect merging \citep[e.g.][]{chambers2001making,quintana2002terrestrial,raymond2009building,izidoro2016asteroid}. 
However, collisional outcomes during planet accretion are diverse and include graze-and-merge, hit-and-run, erosion and catastrophic disruption \citep{leinhardt2012,leinhardt2015, genda2011merging}. 
Recently, studies have modeled planet accretion in simulations that take into account imperfect merging and collisional fragmentation \citep{chambers2013late,quintana2016frequency}. To obtain more information about collisional ejecta, collisional fragmentation and re-accretion of escaped material was investigated in recent $N$-body simulations of planet formation where the outcomes of collisions between bodies were tracked during dynamically excited planet formation such as in the Grand Tack model \citep{carter2015compositional}. The Grand Tack is a model of Jupiter's evolution that was developed to explain the low mass of Mars. In this model, Jupiter migrates inwards and then back outwards to near its current orbit, dynamically scattering mass out of the Mars and asteroid belt region \citep{walsh2011low}. In these dynamically excited simulations, the majority of collisions during and immediately after the tack disrupt the projectile or erode the target. Even in calmer planet formation scenarios without giant planet migration, a significant fraction of collisions are erosive \citep{carter2015compositional}. {Here, we examine the velocities of these collisions to determine how many reach the criteria for impact-induced melting and vaporization.}

The total mass of material that is ejected during impacts and subsequently re-accreted can be a significant fraction of the mass of the final bodies \citep{bonsor2015collisional}. The effects of subsequent re-accretion of ejecta on planets has been studied: \citet{carter2015compositional} examined the {bulk} compositional effects on mantles and cores of growing planets, and \citet{carter2018collisional} investigated crustal erosion, but neither scrutinized the thermodynamic path of {the ejected material.} We know that impacts are capable of melting and vaporizing silicates, but we do not know the scale or conditions in which these phase changes occur {during planet formation, nor the potential of changing the dynamics of impact ejecta}.{ If a significant fraction of planetary mass is processed as melted and vaporized ejecta, re-accretion of this thermally-processed material may have a cumulative effect on the chemistry of the final planets.}

In this work, we use the experimentally constrained forsterite principal Hugoniot from \citet{root2018forsterite} and new measurements of the Gr\"uneisen parameter to calculate specific entropy on the forsterite principal Hugoniot. We use specific entropy on the forsterite and silica Hugoniots to predict melt and vapor production induced by impacts. We then apply these predictions to $N$-body planet formation models to investigate the occurrence of impact-induced melting and vaporization during planet formation.

\begin{table*}
\centering
 \small
 \begin{tabular*}{\textwidth}{@{\extracolsep{\fill}} l l l } 
 \hline
  Symbol & Description & Units   \\ [0.5ex] 
 \hline
 $u_s$ & Shock Velocity & km/s\\
 $u_p$ & Particle Velocity & km/s\\ 
 $T$ & Temperature & K\\ 
 $P$ & Pressure & Pa\\ 
 $S$ & Specific Entropy & J/K/kg\\ 
 $E$ & Specific Internal Energy & J/kg \\

 $C_x$& Specific Heat Capacity at constant $x$ & J/K/kg\\ 
 $\rho$ & Density & kg/m${^{3}}$\\ 
 $V$ & Specific Volume & m$^3$/kg\\ 
 $f$ & Eulerian Strain & ---\\ 
 $\gamma$ & Gr{\"u}neisen Parameter & ---\\ 
 V$_i$ & Impact Velocity & km/s\\ 
 \hline
  \\

\end{tabular*}
\caption{Summary of variables used in this work. Specific heat capacity is typically measured at constant V or constant P.} \label{tab:symbols}
\end{table*}

\begin{table*}
\centering
 \small
 \begin{tabular*}{\textwidth}{ @{\extracolsep{\fill}} l l l l l} 
  \hline
 Parameter & Description & Value & Units & Reference  \\ [0.5ex] 

 \hline
  $\rho_0$ & Initial Liquid Density (1 bar and 3000 K) & $2597(\pm11)$ &kg/m$^{3}$ & \citet{thomas2013direct}\\ 
  $\gamma_0$ & Gr{\"u}neisen Parameter at $\rho_0$ & 0.396& --- & \citet{thomas2013direct}\\ 
  $m_a$ & Molar Mass & $140.6931$ &mol/g & ---\\ 

   $T_m$ & Melting Temperature & $2174$ &K& \citet{richet1993melting}\\ 
   $T_{ref}$ & Isentrope Foot Temperature & $3000$& K& ---\\ 
    $C_V$ & Liquid Isochoric Heat Capacity & $1737.36$ &J/K/kg& \citet{thomas2013direct}\\
    $C_P$ & Liquid Iosbaric Heat Capacity & $1926.18$ &J/K/kg& \citet{thomas2013direct}\\
    $\gamma_{\infty}$ & Infinite Compression Limit of the $\gamma$ & 2/3 & & --- \\[0.5ex] 
    \hline
      \\
      
\end{tabular*}
\caption{Thermodynamic parameters for Mg$_2$SiO$_4$ used in the derivation of specific entropy on the principal Hugoniot. Where uncertainties are not reported in the primary reference, we describe assumed uncertainties in text. }
\label{tab:params}
\end{table*}

\begin{table*}
\centering
 \small
 \begin{tabular*}{\textwidth}{@{\extracolsep{\fill}} l l l }

Material & Hugoniot Equations & Reference \\
\hline 
 Forsterite & $u_s(u_p) = 4.632 + 1.455 u_p + (4.291E{-03}) u_p^2 - (7.844E{-04}) u_p^3$ & Refit from \citet{root2018forsterite}\\ [0.5ex] 
  & $T(u_s)=-183.188u_s + 15.605u_s^2 + 2.785 u_s^3$& \citet{root2018forsterite} \\ [0.5ex] 
  & $S(P)=-28691P^{-0.5}+ 152.8P^{0.5} - 0.0208P^{1.5} + 3680$& This Work \\ 
    & $\gamma(\rho) = \gamma_{\infty} + (0.377 - \gamma_{\infty}) \left( \frac{\rho_0}{\rho} \right)^{3.705} + 0.657 e^{-(\rho - 4930)^2/1192^2}$& This Work \\ 
 \hline
  $\alpha$-quartz & $u_s(u_p) = 1.754 + 1.862 u_p - (3.364E{-02}) u_p^2 + (5.666E{-04}) u_p^3$ & \citet{knudson2013adiabatic}\\ [0.5ex] 
& $S(P)=2820(\pm 106)P^{-0.5}+468.3(\pm1.8)P^{0.5}-6.68(\pm.037)P-593(\pm25.2)$ & Refit from \citet{kraus2012shock}\\
 \hline
   Fused Silica & $u_s(u_p) = 1.386 + 1.647 u_p - (1.146E{-02}) u_p^2 - (0.6952E{-04}) u_p^3$ & \citet{root2019FS}\\ [0.5ex] 
 \hline
    TPX & $u_s(u_p) = 1.795(\pm.018) + 1.357(\pm.003) u_p - 0.694(\pm.027) u_p e^{-0.273(\pm.011) u_p}$ & \citet{root2015shock}\\ [0.5ex] 
 \hline
All  & $\frac{\rho}{\rho_0}=\frac{u_s}{u_s - u_p}$& Conservation of mass\\
   & $P=\rho_0 u_s u_p$& Conservation of momentum\\
   & $E-E_0 = \frac{1}{2}(P+P_0)\left(\frac{1}{\rho_0}-\frac{1}{\rho}\right)$& Conservation of energy\\
  \hline
  \\

\end{tabular*}
\caption{ Equations for the principal Hugoniots of different materials used in this work. Errors on the parameters for equations {$S(P)$} for $\alpha$-quartz, and {$u_s(u_p)$} for {TPX} are given. {Uncertainty for $\gamma(\rho)$ of forsterite is 32\%.} Uncertainty for the others have covariance matrices included in the supplemental materials in Table \ref{tab:fo_fit}. The forsterite principal Hugoniot equations have a range of validity from 200 to 950 GPa, and the $\gamma$ function has a range from 2597 to 6500 kg/m$^3$. The initial specific entropy is 669 J/K/kg, the initial temperature is 298.15 K, and the initial density ($\rho_0$) is 3220 kg/m$^3$. The quartz principal Hugoniot {$u_s(u_p)$} equations have a range of validity 40 to 800 GPa, while the {$S(P)$} equation has a validity between 110 to 800 GPa. The initial specific entropy is 660 J/K/kg, the initial temperature is 298.15 K, and the initial density is 2651 kg/m$^3$. Fused Silica principal Hugoniot equations have a range of validity 200 to 1600 GPa and the initial density is 2200 kg/m$^3$. TPX principal Hugoniot equations have a range of validity up to 985 GPa and an initial density of 833 kg/m$^3$. The material-specific $u_s(u_p)$ equations can be transformed to pressure, specific volume and specific internal energy via the Rankine-Hugoniot conservation equations.}
\label{tab:equations}
\end{table*}

\section{Absolute Entropy on the Forsterite Hugoniot}

\begin{figure}
\centering
\includegraphics[width=\columnwidth]{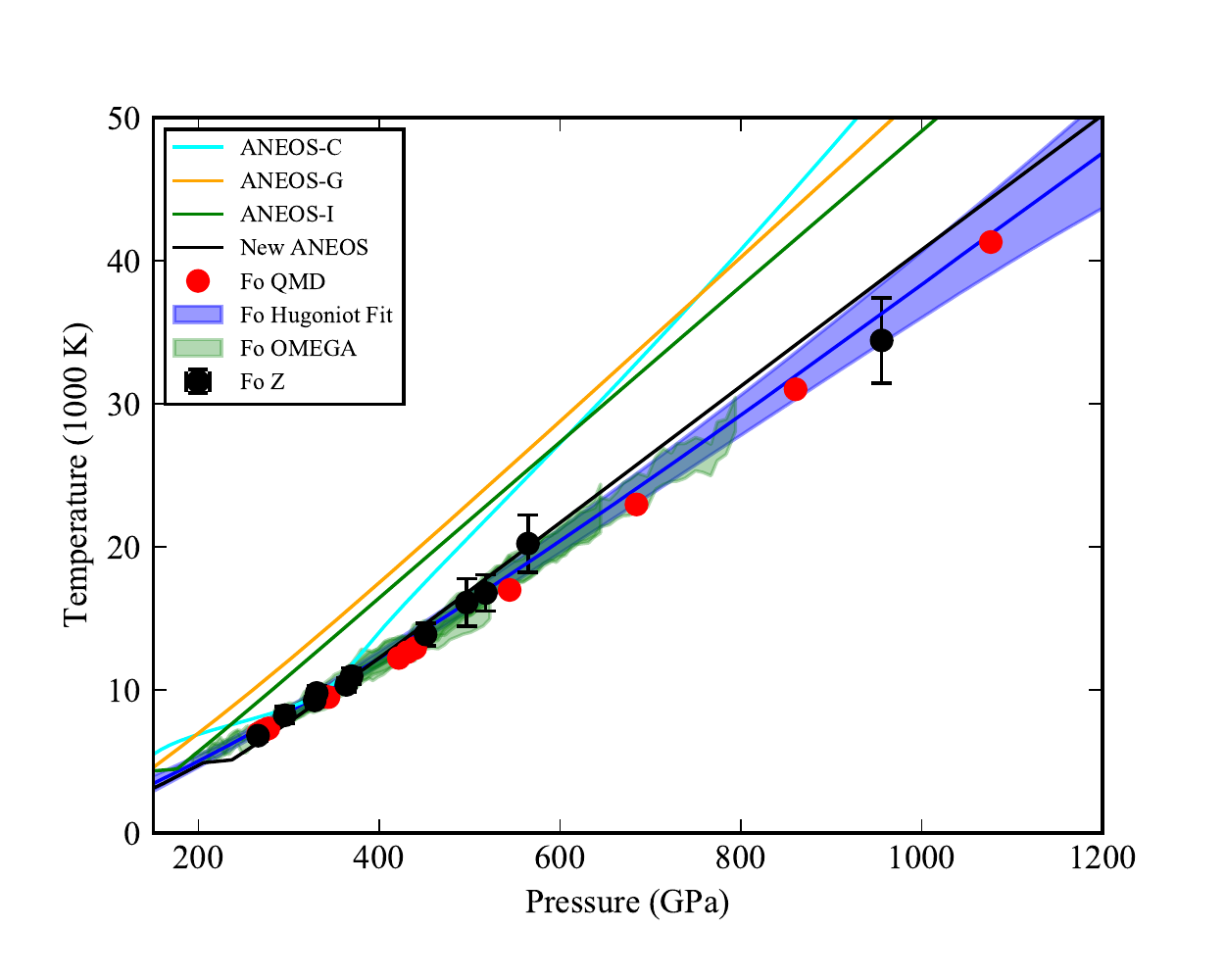}

\caption{The principal shock Hugoniot of forsterite in pressure-temperature space. Red points are QMD calculations, Z Machine data are in black, and the green bands are from decaying shock experiments on the OMEGA laser. The blue line is a fit through the data, with the blue band representing the $1\sigma$ error envelope. QMD, Z, and Omega data are from \citet{root2018forsterite}. Z temperature measurements at about 500 and 560~GPa have been revised and are discussed in Section \ref{ForsteriteDatUp}. The cyan (ANEOS-C \citep{canup2013lunar}), green (ANEOS-I \citep{collins2014improvements}), and orange (ANEOS-G \citep{nakajima2014investigation, cuk2012making}) lines are different forsterite input parameter sets for the ANEOS model \citep{melosh2007hydrocode}. They all over predict the temperature over the entire measured range. The black line presents a revised ANEOS model formulation and forsterite parameter set developed using the newly available data \citep{stewart2019improvements}.} \label{fig:TP}
\end{figure}

The principal Hugoniot for forsterite, shown in Figure~\ref{fig:TP}, was measured up to 950 GPa on two platforms: Sandia National Laboratory's Z Machine and the OMEGA Laser at the University of Rochester \citep{root2018forsterite}. The Z Machine and the OMEGA laser can reproduce most impact energies that are achieved during accretion.
 
The Z Machine is a pulsed power source that is capable of delivering 20 MA of current to the target assemblage over pulses of a few 100 ns \citep{spielman1996pbfa,matzen1997z,savage2007overview}. The current pulse is tailored to accelerate flyer plates while the impact side of the plate remains at a constant density, generating a planar shock wave upon impact with the target \citep{lemke2003characterization,lemke2005magnetically,lemke2011magnetically}. The OMEGA laser facility generates decaying shock waves in the mineral sample by laser-{ablation} of a driver material. By carefully tuning the laser pulse shape \citep{boehly1994upgrade}, thermal emission over a  large, continuous range of shock pressures can be measured in a single experiment. These experiments measured shock velocity and thermal emission using a line VISAR \citep{celliers2004line} and a streaked optical pyrometer \citep{miller2007streaked}. For more details of these experiments, refer to \citet{root2018forsterite}.

The principal Hugoniot data and fitted Hugoniot for forsterite are shown in Figure \ref{fig:TP}. There is excellent agreement between the two experimental platforms and density functional theory based quantum molecular dynamics (QMD) calculations that were also presented in \citet{root2018forsterite}. Two corrected Z temperature measurements are described in Section~\ref{ForsteriteDatUp}. From these measurements, phase changes can be calculated upon isentropic release of shocked material to the reference pressure using the entropy method, as described in section 1. However, specific entropy cannot be measured directly and must be calculated from the thermodynamics of the system. Tables \ref{tab:symbols}, \ref{tab:params}, and \ref{tab:equations} define the symbols, parameters, and Hugoniot equations used in this work, respectively.

\begin{figure}[h]
\centering
\includegraphics[width=\columnwidth]{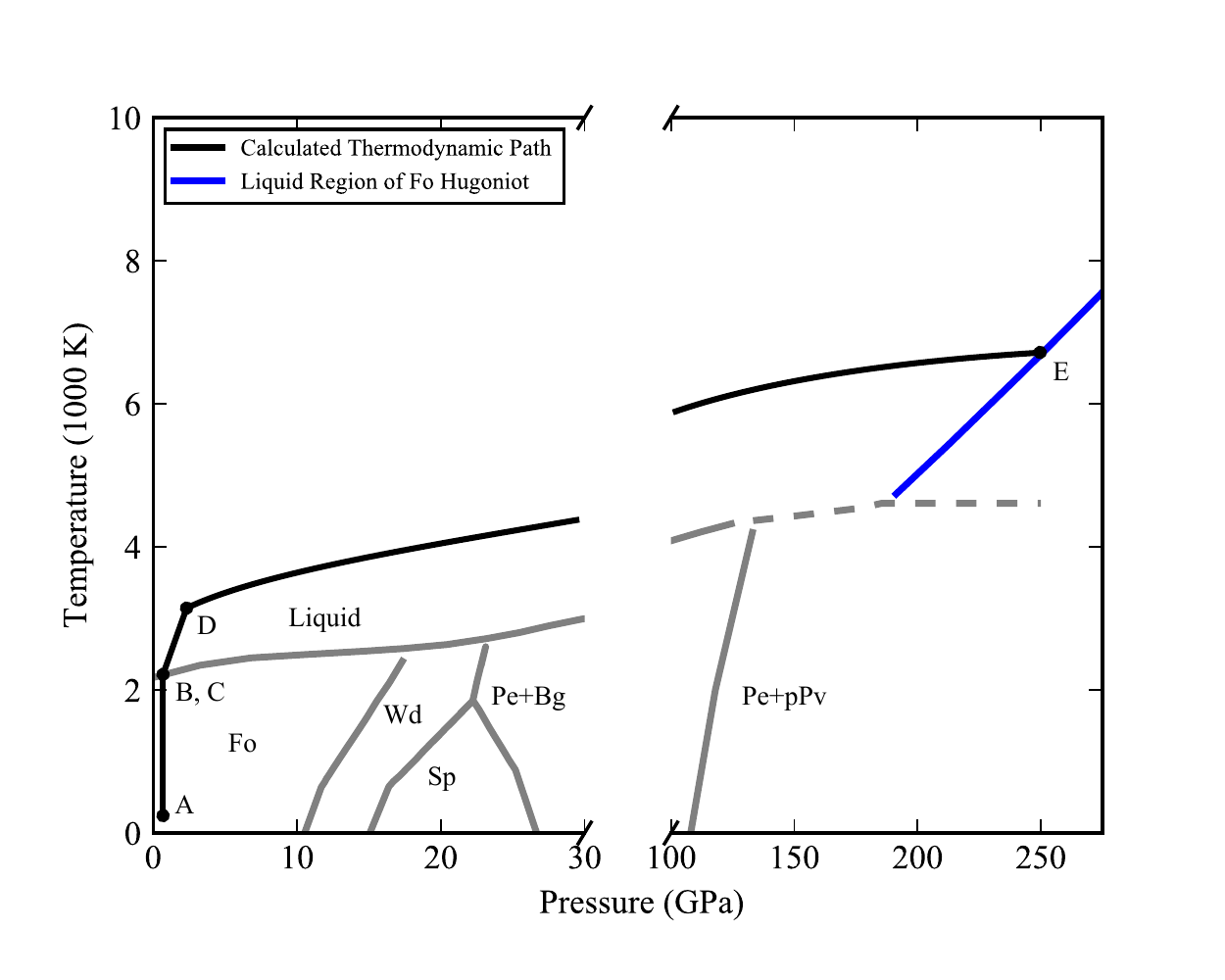}

\caption{Simplified phase diagram of Mg$_2$SiO$_4$ and schematic thermodynamic path used to calculate specific entropy on the principal Hugoniot. Solid phases are forsterite (Fo), wadsleyite (Wd), spinel (Sp), and periclase (Pe) plus bridgmanite {(Bg)}, from \citet{presnall1993melting}. { The postperovskite (pPv) plus periclase field is taken from} \citet{belonoshko2005high}. The melt curve (with dashed extrapolation) is an estimate from \citet{mosenfelder2009mgsio3}. The black line shows the thermodynamic path chosen to integrate for specific entropy, with labels indicating points A through E. The isentrope from point D to E was chosen because it is likely to be completely in the melt region of the Mg$_2$SiO$_4$ system.}
\label{fig:Path_schem}
\end{figure}

To tie the known specific entropy of forsterite at standard temperature and pressure (STP) \citep{robie1982heat} to the measured principal shock Hugoniot, we calculated a thermodynamic integral using available thermodynamic data. Because state variables are independent of the thermodynamic path, the specific entropy on the principal Hugoniot can be calculated using any thermodynamic path. A schematic of the chosen thermodynamic path is shown in Figure \ref{fig:Path_schem}. The path to the principal Hugoniot is separated into 4 steps, bounded by points labeled A through E. Point A is the ambient condition at STP; point B is the state after isobaric heating to the melting point of forsterite; point C accounts for the specific entropy of melting; point D is the state after isochoric heating to the foot of an isentrope; and point E is where this isentrope intersects the principal Hugoniot. Our selected path from point C to point D is in the liquid region and does not intersect the melt curve. While published thermodynamic data for forsterite exists and isentropes in the melt have been calculated \citep{de2008thermodynamics, thomas2013direct, asimow2018melts}, these isentropes are not experimentally constrained above 200 GPa. To be able to calculate isentropes at higher pressures and densities in the liquid region, we performed shallow release experiments concurrently with the shock Hugoniot measurements described above.

\subsection{Shallow Release Experiments to Determine the Gr\"uneisen parameter}

The Gr\"uneisen parameter, $\gamma$, is a thermodynamic parameter needed to calculate thermodynamic states off of the principal Hugoniot. The so-called Mie-Gr\"uneisen approximation assumes that $\gamma$ only depends on volume such that, 
\begin{linenomath*}
\begin{equation}
    \gamma = V \left( \frac{dP_{\rm th}}{dE_{\rm th}}\right)_V,
    \label{Gamma_Defined}
\end{equation}
\end{linenomath*}
where the $dP_{\rm th}$ and $dE_{\rm th}$ are the thermal pressure and thermal specific internal energy.
Furthermore, the thermodynamic relation, 
\begin{linenomath*}
\begin{equation}
\gamma= - \left( \frac{d \ln \, T}{d \ln \, V} \right)_S,
\label{isentrope_gamma}
\end{equation}
\end{linenomath*}
provides temperature along an isentrope.
Thus $\gamma$ is the critical parameter that describes the intersection of the isentrope with the principal Hugoniot (points D to E {in Figure} \ref{fig:Path_schem}).

In most solids, during compression, $\gamma$ decreases with decreasing specific volume such that the following empirical relation is often used,
\begin{equation}
    \gamma = \gamma_0 \left( \frac{V}{V_0} \right)^{q},
\end{equation}
where $q\sim1$ for many materials \citep{asimow2018melts}. At high compression, $\gamma$ must approach a limiting value, which is often taken to be either 2/3, for a free electron gas, or 1/2, for the Thomas-Fermi limit \cite[see discussion in][]{burakovsky2004analytic}. At STP, $\gamma_0=1.29(\pm1)$ for solid forsterite \citep{gillet1991high}.

In contrast to solids, the $\gamma$ for liquid silicates are known to increase with compression \citep{de2008thermodynamics,thomas2013direct, asimow2018melts}. The $\gamma$ for initially 2273~K liquid forsterite has been inferred from shock experiments up to 114.3~GPa, with $\gamma=0.396$ at ambient density and increasing to 1.3 at a density of 4.68~g/cm$^{3}$ \citep{thomas2013direct}. There is currently no data for the $\gamma$ of liquid forsterite at higher compression. The value must decrease at the high-compression limit, so functional forms fit to lower pressures cannot be extrapolated. Here, we determine the Gr\"uneisen parameter for liquid forsterite at high compression using shallow release experiments that constrain isentropic paths in $P-\rho$ space.

Shallow release experiments measure the $P-\rho$ change of release isentropes upon decompression from the shocked state.
These experiments are performed by backing the forsterite sample with a transparent and lower-impedance material. The shock wave compresses the sample and transitions into the backing window that has a known Hugoniot. At the interface, the shock releases by isentropic decompression to the impedance of the window and a lower-pressure shock wave propagates forward into the window and a rarefaction wave partially decompresses the sample. At the material interface, the Rankine-Hugoniot conservation conditions are satisfied so $P$ and $u_p$ are identical in both materials. $P$ and $u_p$ are calculated from the measured $u_s$ using known shock Hugoniots. The only measurement needed to determine the shocked state in the window, and therefore the partial release state in the forsterite sample, is $u_s$. For these experiments, three window materials were used, $\alpha$-quartz, fused silica, and polymethylpentene, another commonly used standard window referred to as TPX, that have known principal shock Hugoniots \citep{knudson2013adiabatic, root2019FS,root2015shock}. The experimental configuration was otherwise similar to \citet{root2018forsterite} to generate planar shock experiments at Z. The measured shock states in forsterite and the partial release states at the windows are given in Table \ref{tab:Volume_gamma}.

\begin{figure*}
\centering
\includegraphics[width=5in]{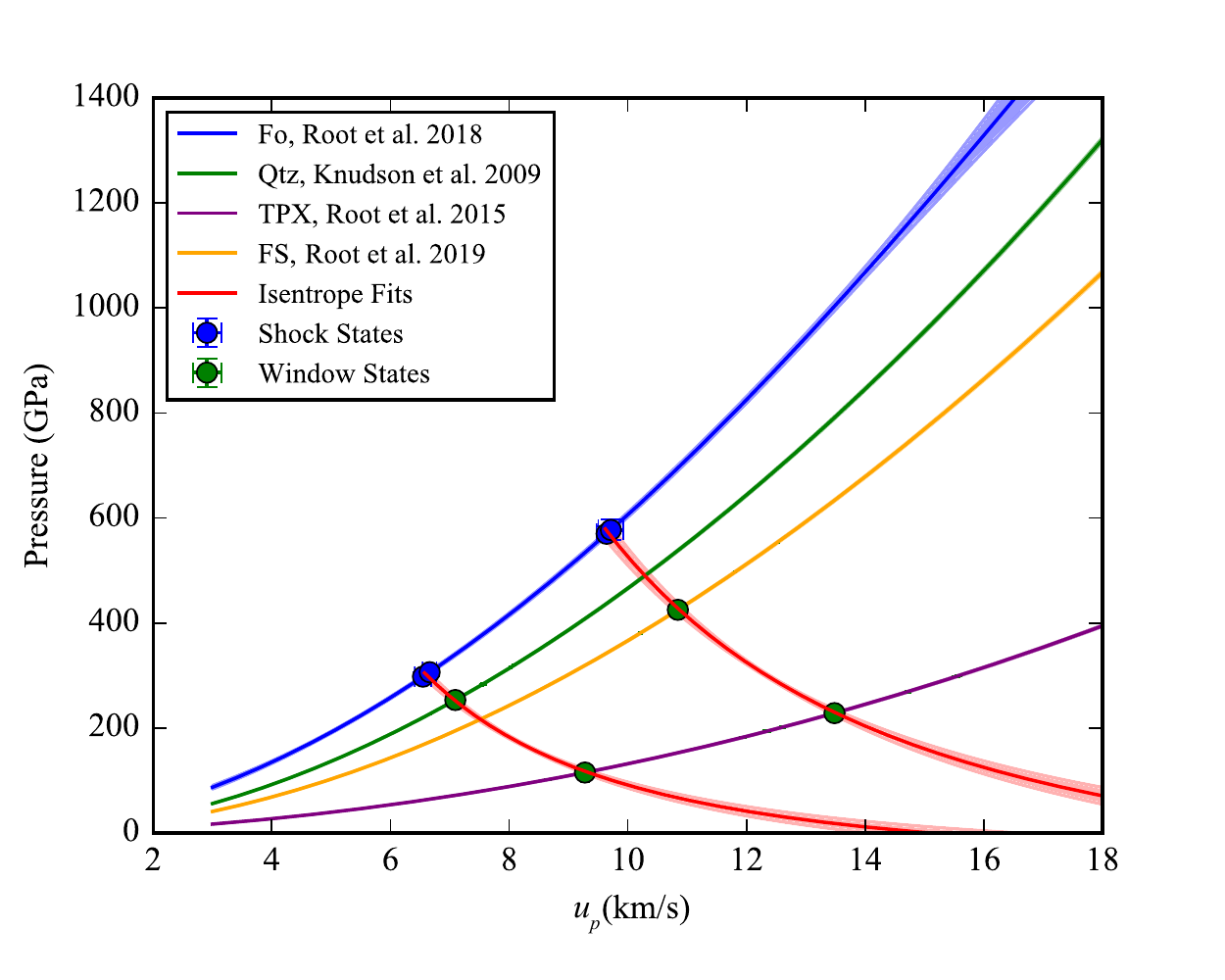}
\includegraphics[width=5in]{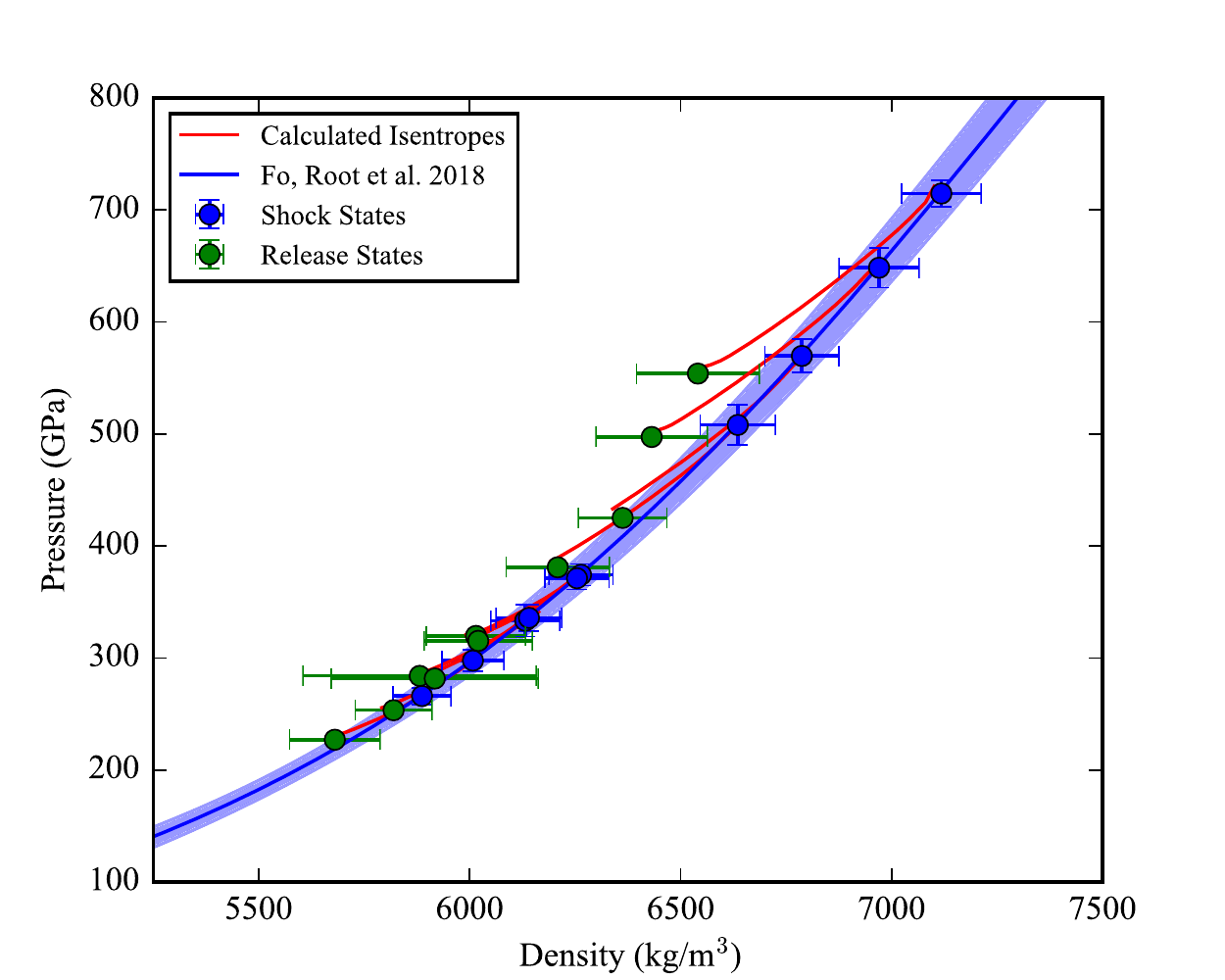}

\caption{ Results from shallow release experiments at the Z machine. (Top) Example fitted release isentropes with uncertainty envelopes (red) shown between the principal Hugoniots of forsterite (Fo), quartz (Qtz), TPX, and fused silica (FS) in pressure-particle velocity space. We derive the densities where the isentropes intersect the partial release states. Residuals of states from the $P$-$u_p$ fit are incorporated into the uncertainty of the calculated density. (Bottom) All shallow release states used to calculate the Gr\"uneisen parameter for liquid forsterite and calculated release isentropes (red).}
\label{fig:P_up_fit}
\end{figure*}

The change in volume during isentropic decompression is given by the Riemann integral, 
\begin{linenomath*}
\begin{equation}
V_r = V_H + \int_{u_{pr}}^{u_{pH}} \left( \frac{du_{p}}{dP} \right)_S du_{p},
\label{eq:riemann}
\end{equation}
\end{linenomath*}
where subscript $r$ is the released state, subscript $H$ is the shock Hugoniot state, and subscript $S$ denotes constant specific entropy \citep{rice1958compression}. This integral is valid only for isentropic processes. Equation \ref{eq:riemann} requires a fitted curve between the shock and release states to evaluate the integral. For this work, we found that the function $P=A u_p^{-2} + B$, where $A$ and $B$ are fitting parameters, could fit the release path data and be simple to evaluate in the integral above. Experiments with identical shock states are grouped and fit together to measure multiple points on the same isentrope. An example of such a fit and calculated volumes are shown in Figure \ref{fig:P_up_fit}.
 
To obtain $\gamma$, we calculate an isentrope that connects the shocked state to the released state by a Mie-Gr{\"u}neisen release isentrope. The isentrope requires a reference curve with known $P$-$V$-$E$ and we reference the forsterite principal Hugoniot in a stepwise fashion as in \citet{mcqueen1970equation}. Because the density of the release states in TPX are below the range of validity of the reference curve, they are not considered to calculate $\gamma$. Pressure on the isentrope for each step is calculated via finite difference,
 \begin{linenomath*}
\begin{equation}
P_i = \frac{P_H - \left( E_H - E_{i-1} + P_{i-1}\frac{\Delta V}{2}\right) \left( \frac{ \gamma(\rho)}{V}\right)_i }{1+\frac{\Delta V}{2} \left( \frac{ \gamma(\rho)}{V}\right) _i},
\label{eq:isentrope}
\end{equation}
\end{linenomath*}
where the specific internal energy is
\begin{linenomath*}
\begin{equation}
E_i = E_{i-1} - \frac{P_i + P_{i-1}}{2} \Delta V.
\end{equation}
\end{linenomath*}
The subscript $i$ is the current element on the isentrope, indexed with density, subscript $H$ is the reference Hugoniot state at the same density, and $\gamma(\rho)$ is some function of the Gr{\"u}neisen parameter dependent only on density. We examine several different formulations for $\gamma$ to solve equation~\ref{eq:isentrope}: linear, exponential, and Gaussian, which are described in detail in Section \ref{Sup:gammasection}. Because the range of densities between the shock and release states is limited to 5500-7500 kg/m$^{3}$, the formulations are intentionally simple. All formulations include a single fitting parameter, $q$, which is iterated until the pressure and density on the isentrope match the shock and release states.

\begin{figure}[ht]
\centering

\includegraphics[width=\columnwidth]{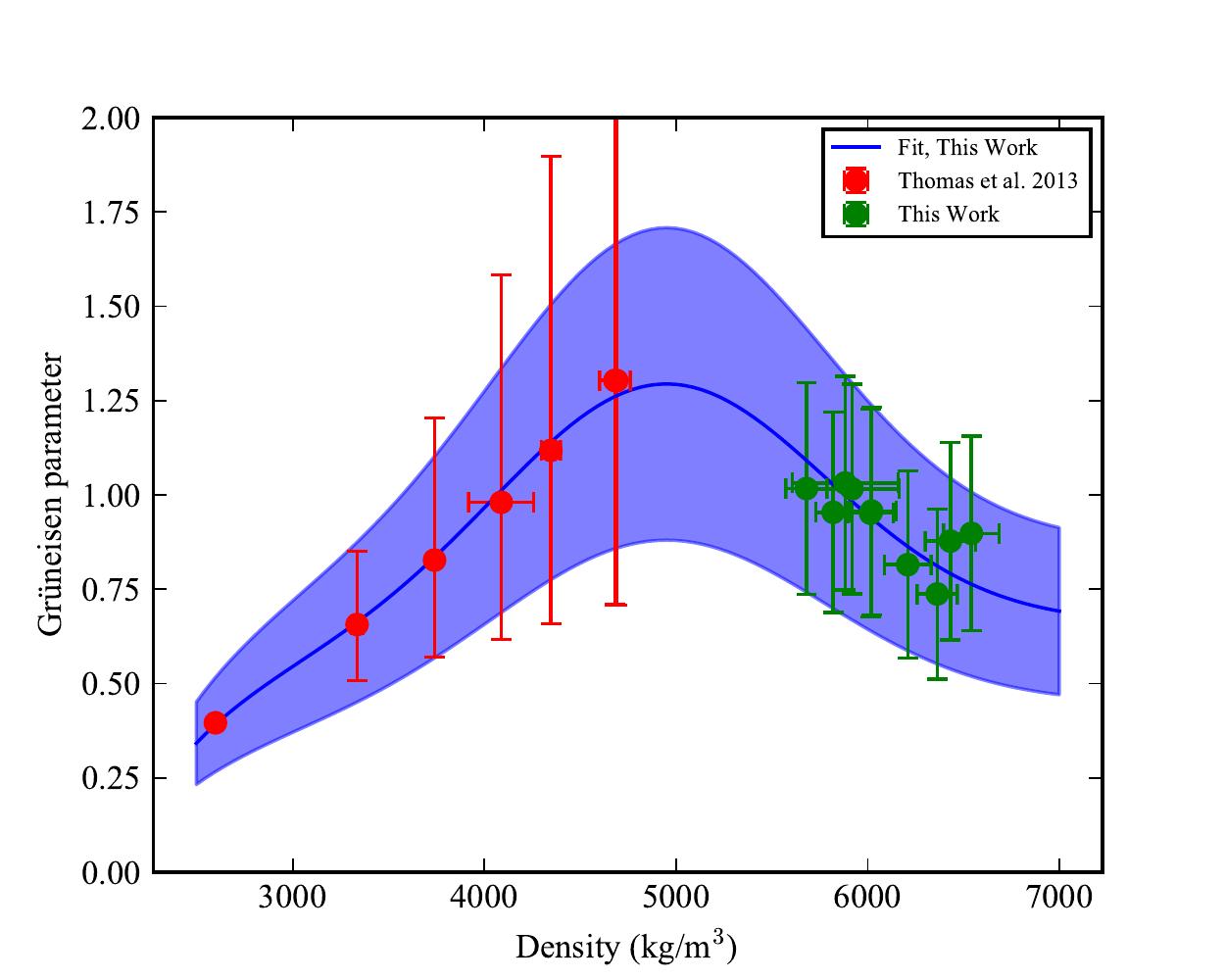}
\caption{Experimentally derived values and our fit to the Gr{\"u}neisen parameter for liquid forsterite.  }
\label{fig:gamma}
\end{figure}

We found that the formulation with the smallest residuals to calculate the isentropic release paths, shown in Figure \ref{fig:P_up_fit}, was the Gaussian formulation (Section \ref{Sup:gammasection}). The calculated values of $\gamma$ are shown in Figure \ref{fig:gamma}. The whole data set is given in Table \ref{tab:Volume_gamma}. 

Our $\gamma$ values were combined with the lower-density data from \citet{thomas2013direct}. We find that the values for $\gamma$ decrease with increasing density over our measured range. The values at the highest densities are consistent with theoretical limits \citep{burakovsky2004analytic}. None of the standard formulations for $\gamma(\rho)$ fit the entire liquid density range. To fit both sets of data, a fit of the form $\gamma(\rho) = \gamma_{\infty} + (A - \gamma_{\infty}) \left( \frac{\rho_0}{\rho} \right)^{B} + C e^{-(\rho - D)^2/E^2}$ is used. Our fit is only valid between the densities of 2597-6500 kg/m$^3$ where data exist. This form is a combination of previously used exponential forms \citep{thomas2013direct} for low densities and the Al\lq tshuler form for high density limits of $\gamma$. All parameters are fit except $\rho_0$, which is the initial liquid density and the infinite compression limit, $\gamma_{\infty}$, which we take as 2/3. A more robust discussion on the mechanics of $\gamma$ is beyond the scope of this study. The fit is given in Table \ref{tab:equations} {with 1-sigma uncertainty estimated at 32\%}. The data and fit are shown in Figure~\ref{fig:gamma}.

\begin{center}
\begin{table*}[ht]
\centering
 \small
 \begin{tabular*}{\textwidth}{@{\extracolsep{\fill}} l l l l l l l l} 
  \hline
 Shot \#  &Win.& V$_i$&Sam. $u_s$&Win. $u_s$ & $\rho_{s}$& $\rho_{r}$ & $\gamma$  \\ 
  & &(km/s) &(km/s)&(km/s)& (kg/m$^3$)& (kg/m$^3$) &  \\[0.5ex] 
 \hline
 Z2792 N7&Qtz &14.82($\pm$.12) &14.75($\pm$.24)& 14.15($\pm$.08)& 6132($\pm$82)& 5882($\pm$276) & 1.03($\pm$.28)\\
 Z2792 S7&Qtz &14.72($\pm$.07) &14.8($\pm$.19)& 14.10($\pm$.06)& 6141($\pm$78)& 5917($\pm$245) & 1.01($\pm$.28)\\ 
 Z2868 N7&Qtz &15.83($\pm$.07) &15.46($\pm$.14)& 14.9($\pm$.06)& 6264($\pm$76)& 6015($\pm$118) & 0.95($\pm$.28)\\ 
 Z2868 S7&Qtz &15.67($\pm$.06) &15.41($\pm$.15)& 14.81($\pm$.05)& 6255($\pm$76)& 6020($\pm$128) & 0.95($\pm$.28)\\ 
 Z2879 N7&Qtz &12.95($\pm$.04) &13.50($\pm$.13)& 12.83($\pm$.04)& 5887($\pm$69)& 5680($\pm$107) & 1.02($\pm$.28)\\ 
 Z2879 S7&Qtz &13.82($\pm$.08) &14.11($\pm$.16)& 13.46($\pm$.06)& 6008($\pm$74)& 5819($\pm$91) & 0.95($\pm$.27)\\ 
 Z3033 N3&FS &18.89($\pm$.09) &17.50($\pm$.24)& 16.94($\pm$.06)& 6636($\pm$89)& 6208($\pm$122) & 0.81($\pm$.25)\\ 
 Z3033 N5&TPX &18.84($\pm$.08) &17.49($\pm$.19)& 19.17($\pm$.05)& 6634($\pm$85)& 5119($\pm$121) & ---\\ 
 Z3033 S3&FS &20.32($\pm$.06) &18.34($\pm$.18)& 17.81($\pm$.04)& 6787($\pm$88)& 6362($\pm$105) & 0.74($\pm$.23)\\ 
 Z3033 S5&TPX &20.37($\pm$.11) &18.44($\pm$.24)& 20.32($\pm$.04)& 6806($\pm$94) & 5190($\pm$133) & ---\\ 
 Z3044 N3&TPX &21.79($\pm$.07) &19.34($\pm$.18)& 21.94($\pm$.05)& 6971($\pm$94)& 5053($\pm$163) & ---\\ 
 Z3044 N5&FS &21.75($\pm$.06) &19.34($\pm$.20)& 19.13($\pm$.04)& 6971($\pm$95)& 6432($\pm$132) & 0.88($\pm$.26)\\ 
 Z3044 S3&TPX &23.2($\pm$.08) &20.11($\pm$.10)& 23.07($\pm$.05)& 7115($\pm$94)& 4105($\pm$115) & ---\\ 
 Z3044 S5&FS &23.2($\pm$.05) &20.13($\pm$.10)& 20.09($\pm$.04)& 7118($\pm$94)& 6541($\pm$145) & 0.90($\pm$.26)\\ 
 Z3101 N1&TPX &12.74($\pm$.08) &13.42($\pm$.11)& 14.09($\pm$.05)& 5871($\pm$67)& 4643($\pm$151) & ---\\ 
 Z3101 S1&TPX &13.83($\pm$.11) &14.27($\pm$.12)& 14.89($\pm$.06)& 6039($\pm$70)& 4819($\pm$132) & ---\\ 
 Z3172 N7&TPX &17.9($\pm$.10) &16.88($\pm$.15)& 18.30($\pm$.06)& 6523($\pm$80)& 5068($\pm$122) & ---\\ 
 Z3172 S1&TPX &19.16($\pm$.06) &17.57($\pm$.16)& 18.83($\pm$.07)& 6647($\pm$83)& 5268($\pm$194) & ---\\ 
 Z3201 S1&TPX &23.14($\pm$.05) &20.10($\pm$.19)& 23.02($\pm$.09)& 7112($\pm$99)& 5119($\pm$142) & ---\\ 
  \hline
\end{tabular*}
\caption{Compiled results of experiments, measurements and calculations for partial release of forsterite at the Z-Machine. All experiments are a forsterite sample backed by a window to tamper the release. Shock velocities in the forsterite samples are corrected from and in the same manner as \citet{root2018forsterite}, based on acceleration of the shock front at the sample-window interface. Densities at the interface before decompression are recalculated based on the corrected shock velocity. $\gamma$'s are given at the density of the release state of forsterite.}
\label{tab:Volume_gamma}

\end{table*}
\end{center}

\subsection{Entropy on the Principal Hugoniot}

We calculated specific entropy change from STP to the principal Hugoniot along the path (A-E) in Figure~\ref{fig:Path_schem}. The total specific entropy is given by
\begin{linenomath*}
\begin{equation}
S_{\rm Total}=S_{\rm STP}+\Delta S_{\rm Solid-Heating}+\Delta S_{\rm Melting}+\Delta S_{\rm Liquid-Heating}.
\end{equation}
\end{linenomath*}
Starting from the specific entropy at STP ($S_{\rm STP}=$ 669(1) J/K/kg, Table \ref{tab:entropy}), we calculated the increase in $S$ due to heating forsterite isobarically to the melting point at 1 bar (point A to B). For heating at constant pressure, the change in specific entropy is given by;
\begin{linenomath*}
\begin{equation}
\Delta S_{\rm Solid-Heating}=\int^{T_{\rm Melting}}_{298.15} \frac{C_P (T)}{T} dT,
\label{eq:Sheat}
\end{equation}
\end{linenomath*}
where $T$ (K) is the temperature and $C_P(T)$ (J/mol/K) is taken from \citet{gillet1991high},
\begin{linenomath*}
\begin{equation}
\begin{multlined}
C_P(T)=-402.753+74.29 \ln(T)+\frac{\rm 87.588E3}{T} \\
 -\frac{\rm 25.913E6}{T^2}+\frac{\rm 25.374E8}{T^3}.
 \end{multlined}
\end{equation}
\end{linenomath*}
The total change in specific entropy from the isobaric heating to 2174 K is $\Delta S_{\rm Solid-Heating}=$ 2339($\pm$195) J/K/kg. The specific entropy associated with melting (point B to C) is taken from \citet{richet1993melting}, where $\Delta S_{\rm Melting}=$ 464($\pm$4.3) J/K/kg. From the melting point, we calculate the specific entropy of  isochorically heating liquid forsterite to a reference isentrope at 3000 K (point C to D). This is a similar integral as before, where the only difference is that volume is held constant instead of pressure, 
\begin{linenomath*}
\begin{equation}
\Delta S_{\rm Liquid-Heating}=\int^{T}_{T_{\rm Melting}} \frac{C_V }{T} dT.
\label{eq:Lheat}
\end{equation}
\end{linenomath*}
The heat capacity at constant volume is taken from \citet{thomas2013direct},  giving $\Delta S_{\rm Liquid-Heating}=$ 559($\pm$80) J/K/kg.

Table \ref{tab:entropy} summarizes the specific entropy at each step, and the total at the base of the isentrope. Now that we have specific entropy at the foot of the isentrope, we calculate the intersection with the principal Hugoniot (point D to E) in temperature and density.

\begin{figure}[h]
\centering
\includegraphics[width=\columnwidth]{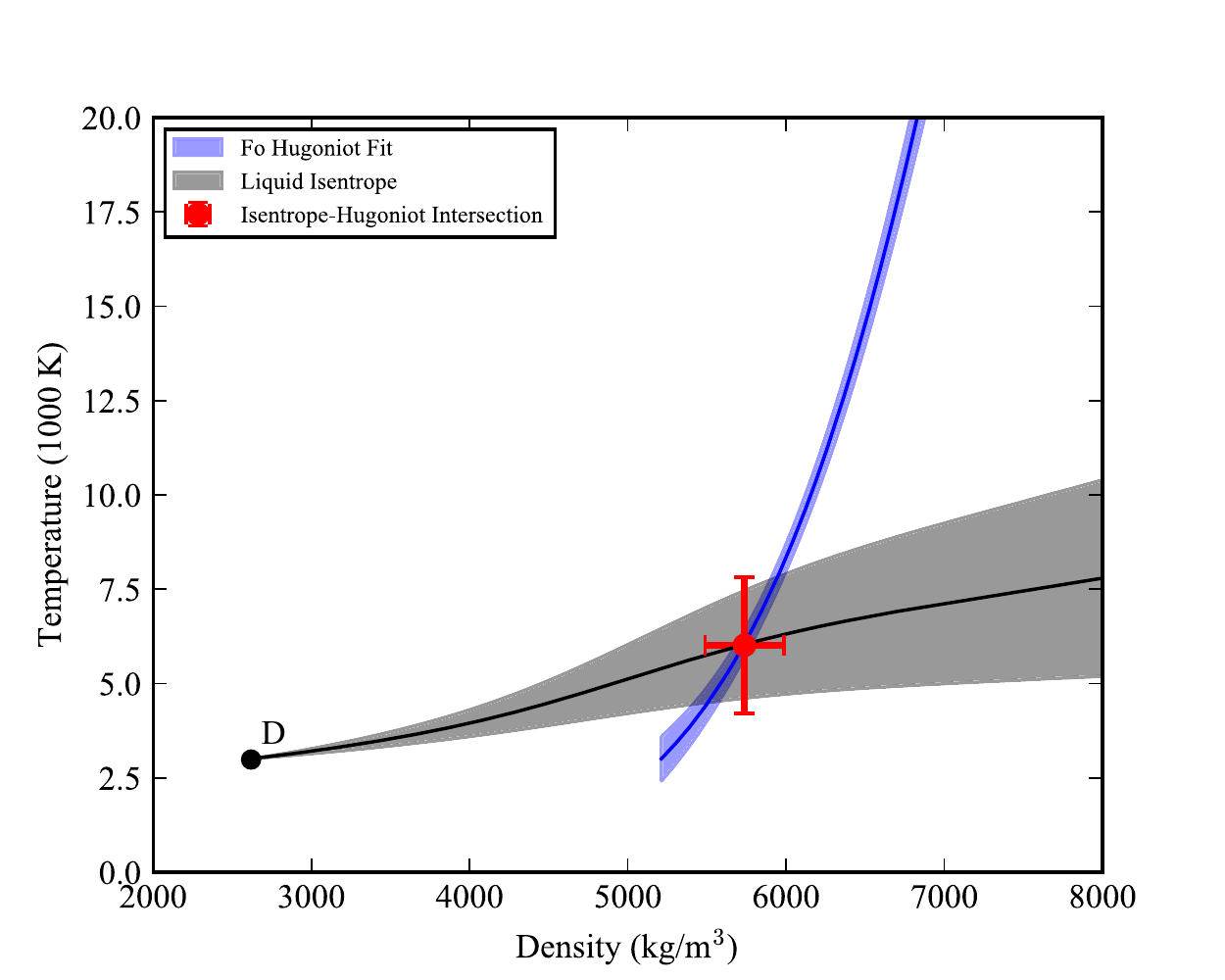}

\caption{In density and temperature space, the principal Hugoniot of forsterite (blue line) and the calculated isentrope (black line) used for relating a known specific entropy at STP to a point along the Hugoniot. The intersection between the principal Hugoniot and isentrope is shown by the red point. Uncertainty on the intersection is dominated by uncertainty of the Gr{\"u}neisen parameter. Point D is the foot of the defined isentrope as in Table \ref{tab:entropy}. }
\label{fig:Path_calc}
\end{figure}

The Gr{\"u}neisen parameter along an isentrope is given in Eq. \ref{isentrope_gamma}.
Approximating infinitesimal steps along the isentrope and substituting specific volume with density, the temperature on the isentrope is given stepwise by
\begin{linenomath*}
\begin{equation}
T_i = \exp \left( \gamma(\rho_i) [\ln \, \rho_i-\ln \, \rho_{i-1}]+\ln \, T_{i-1} \right),
\end{equation}
\end{linenomath*}
where $\gamma$ is defined in Table \ref{tab:equations}. The liquid forsterite is compressed along this isentrope until it intersects the principal shock Hugoniot, giving the specific entropy at one pressure-temperature-density point on the Hugoniot. The intersection and calculated path are shown in Figure \ref{fig:Path_calc}. The thermodynamic values of the intersection and each step of the thermodynamic integral are summarized in Table \ref{tab:entropy}.

\begin{center}
\begin{table*}
\centering
 \small
 \begin{tabular*}{\textwidth}{@{\extracolsep{\fill}} l l l l l l l l } 
  \hline
 Point  & Description & $P$ (GPa) &$\rho$ (kg/m$^3$) & $T$ (K) & $S$ (J/K/kg) & $\Delta S$& Entropy Reference  \\ [0.5ex] 
 \hline
 A & STP &$10^{-4}$&$3220(\pm9.7)$&298.15& $669(\pm1)$& ---& \citet{robie1982heat}; \\ 
 & & & & & & &\citet{robie1995thermodynamic}\\ 

 B  & Isobaric heating &$10^{-4}$&$2995(\pm10)$*&$2174(\pm100)$& $3008(\pm195)$&$2339(\pm195)$& Equation \ref{eq:Sheat}\\ 

 C & Melting &$10^{-4}$&$2597(11)$&$2174(100)$& $3474(\pm195)$&$464(\pm4.3)$& \citet{richet1993melting}\\ 

 D & Isochoric heating &$2.3(\pm.4)$&$2597(\pm11)$&$3000$& $4033(\pm211)$&$559(\pm80)$& Equation \ref{eq:Lheat}\\ 

 E  & Isentropic compression &$231(\pm54)$&$5737(\pm249)$&$6013(\pm1803)$& $4033(\pm211)$ & ---& \\
  \hline
\end{tabular*}
\caption{ Pressure, density, temperature, and specific entropy values for each step of the thermodynamic integral from STP (A), isobaric heating to the melting point (B), melting (C), isochoric heating to 3000 K (D), to an intersection point on the principal Hugoniot (E). *density prior to melting is from \citet{bouhifd1996thermal}.}
\label{tab:entropy}

\end{table*}
\end{center}

Once specific entropy is known at one point on the principal Hugoniot, the specific entropy along the principal Hugoniot can be found via the first {and second }law{s} of thermodynamics, 
\begin{equation}
dS=\frac{dE}{T}+\frac{PdV}{T} \label{firstlaw}
\end{equation}
where $dS$ is the differential specific entropy and $dV$ is the differential specific volume. $dE$, the differential specific internal energy is known by the Rankine-Hugoniot conditions, and temperature on the principal Hugoniot is known between 200 and 950 GPa \citep{root2018forsterite}. We fit a polynomial to our derived values for specific entropy on the principal Hugoniot, which is presented in Figure~\ref{fig:t-s} and Table~\ref{tab:equations}.  

Uncertainty is propagated through all calculations using a \textit{Monte-Carlo} uncertainty analysis technique. All variables with a measured uncertainty are randomly perturbed according to a normal distribution about their 1-sigma uncertainty. The calculations are repeated with such random perturbations until the resulting data cloud converges to a Gaussian, typically 10\,000 steps. The mean of the Gaussian is taken as the fitted value, and the standard deviation to be the one sigma uncertainty. For this work, the forsterite principal Hugoniot was refit using a cubic polynomial for the purposes of facilitating \textit{Monte-Carlo} uncertainty analysis. We refit using data from \citep{root2018forsterite} and available gas gun data above $u_p > 4$ km/s \citep{mosenfelder2007thermodynamic,lyzenga1980shock,jackson1979shock,watt1983shock}.
The largest source of uncertainty in this work is the uncertainty on the Gr{\"u}neisen parameter, and future work should focus on decreasing the uncertainty between 4000 and 5000 kg/m$^3$.

\subsection{Comparison to Hydrocode Models for Forsterite}

\begin{figure*}
    
    \centering
    \includegraphics[width=5in]{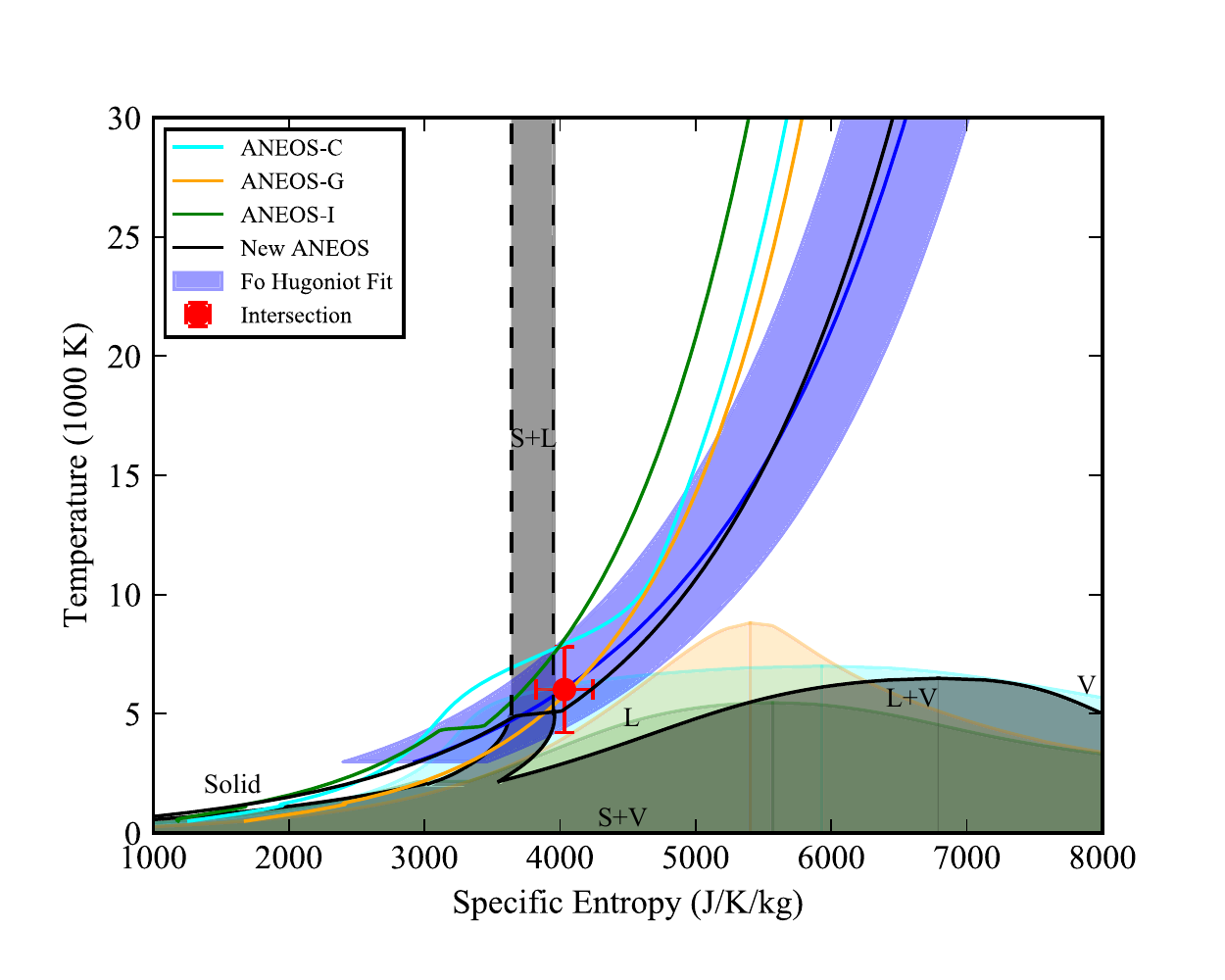}
    \includegraphics[width=5in]{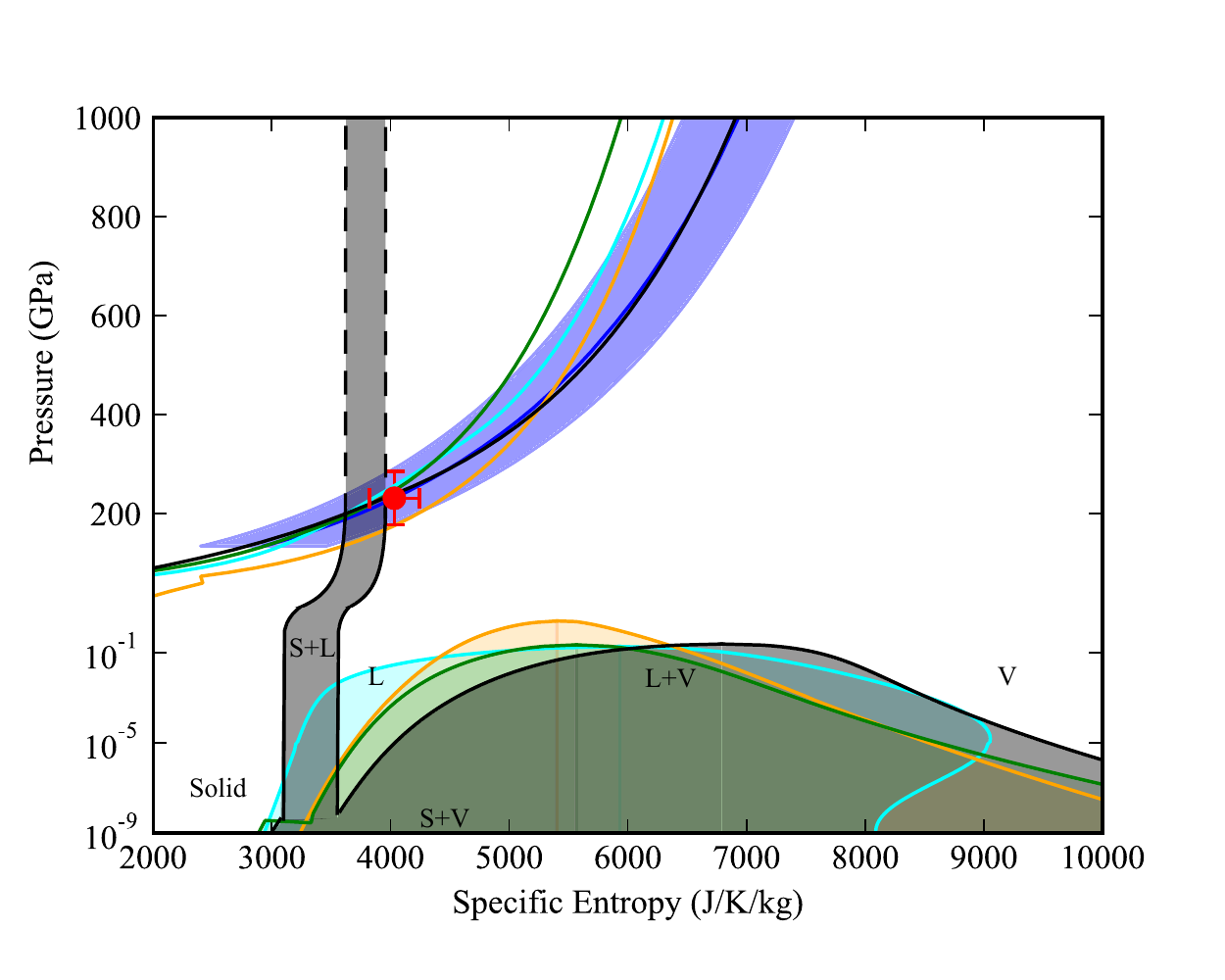}
    \caption{The principal Hugoniot is shown in blue with its associated error envelope in specific entropy space. Other solid lines are the predicted Hugoniot and liquid-vapor dome from the various ANEOS parameterizations. The red point is the calculated intersection between the isentrope and Hugoniot. The dashed black lines represent a schematic of the melt curve. (Top) Temperature against specific entropy and (Bottom) Pressure against specific entropy of the principal Hugoniot and the ANEOS models. The divergence between the previously predicted and the measured Hugoniots should not be ignored, however, the models do lie within the extremes of the uncertainty. The new ANEOS model Hugoniot reproduces the forsterite principal Hugoniot better than the previous models. } 
    \label{fig:t-s}
\end{figure*}

Our new thermodynamic data on the liquid region of the forsterite EOS provide new constraints for development of revised EOS models. 
Our newly calculated principal Hugoniot is compared to different model Hugoniots in Figure~\ref{fig:t-s}. Here, we refer to three different versions of the ANEOS model for forsterite as ANEOS-C \citep{canup2013lunar}, ANEOS-I \citep{collins2014improvements}, and ANEOS-G \citep{cuk2012making,nakajima2014investigation}. These model parameter sets were developed before the availability of the high pressure data used in this work. All previous ANEOS models consistently over predict temperature in shocks above 200 GPa, with greater divergence in specific entropy at higher pressures. \citet{root2018forsterite} showed that the model principal Hugoniot from ANEOS-G falls within the experimental uncertainties in pressure-density space. Further comparisons are made between the new high pressure data and ANEOS models in \citet{stewart2019improvements}. 

From the Rankine-Hugoniot conservation equations, the change in specific internal energy between shock states is only dependent on changes in density and pressure. Given the first law of thermodynamics, if the change in temperature is over predicted then the change in entropy is off as well. The new ANEOS model improves the fit in the liquid region of the Hugoniot because of the implementation of adjustable limit to the specific heat capacity. The new model parameters were focused on fitting the liquid and vapor regions of the phase diagram, see \citep{stewart2019improvements}.

\section{Shock-Induced Phase Changes}

Using our new calculations of specific entropy of shocked forsterite, we revisit the question of the onset of melting and vaporization during planetary collisions. The criteria for the onset of a phase change depends on the ambient pressure and initial conditions. For applications to planetary collisions, we use two ambient reference pressures: 1 bar and the triple point. Example isentropic release paths that decompress to these pressures are shown in Figure \ref{fig:ent_schem}.

\citet{kraus2012shock} calculated the specific entropy on the silica principal Hugoniot. Here, we have extended the entropies to lower pressures and revised the fit (See Section \ref{Sup:quartzSection}). We present revised criteria for shock-induced phase changes of silica, a less refractory phase, for comparison to our new forsterite results.

\subsection{Reference States and Initial Conditions}
To use the entropy method, we require the specific entropies of phase boundaries at the ambient reference pressures: 1 bar and the triple points for forsterite (5.2 Pa, \citet{nagahara1994evaporation}) and $\alpha$-quartz (2 Pa, \citet{mysen1988condensation}).  

For forsterite, the specific entropy of complete melting at 1 bar is taken from our integration, point C in Table \ref{tab:entropy}, and we assume that the specific entropy of complete melting at the triple point is within uncertainty of the 1 bar value. There is no experimental data for the boiling point of forsterite at 1 bar. Here, we estimate the 1-bar boiling temperature  {as} 3300$\pm$300 K. {This is based on evaporation experiments} \citep{nagahara1994evaporation} {extrapolated to 1 bar, 3265 $\pm$ 473 K, the new ANEOS boiling point, 3375 K} \citep{stewart2019improvements}{, and that forsterite's boiling point must be higher than that of silica, 3127 K} \citep{chase1998nist}.  We calculate the specific entropy of incipient vaporization at 1 bar by thermodynamic integral similar to Eq. \ref{eq:Sheat}, using a constant isobaric heat capacity from \citet{thomas2013direct}, from the melting temperature to the boiling point. The specific entropy at incipient vaporization at the triple point is simply the specific entropy at complete melting. Specific entropies at 50\% vaporization are not experimentally constrained and vary greatly {amongst} the model calculations. For this work, we use the model vapor curve from \citet{stewart2019improvements}, labelled new ANEOS in Figure~\ref{fig:t-s}. While this work experimentally constrains the specific entropy of the principal forsterite Hugoniot, the pressure-volume-temperature states at higher pressures on the liquid-vapor dome still need experimental validation. For $\alpha$-quartz, we use the entropies for the phase boundaries given in \citet{kraus2012shock}. The specific entropy of complete melting at the triple point and one bar are assumed to be identical. Table \ref{OnsetTable} presents all of the specific entropies of all of the phase changes considered in this work. 

The interiors of differentiated bodies in the early solar system are generally warmer than room temperature. To illustrate the effects of the initial temperature on the criteria for shock-induced phase changes, we calculate two different initial temperatures for forsterite. The first initial condition in both materials is STP. For warmer initial conditions, we chose an initial temperature of 1200~K for forsterite. For {this} warmer initial condition, {we use the new ANEOS model} \citep{stewart2019improvements}.
1200~K was chosen for forsterite to be near the low-pressure solidus temperature of terrestrial mantle composition. Uncertainties are only calculated for forsterite and $\alpha$-quartz with initial conditions at STP and we have not propagated uncertainties for the warm initial condition{ of forsterite}.

\subsection{Criteria for Phase Changes}

Table \ref{OnsetTable} presents the critical pressures, impact velocities, and entropies to reach melting and vaporization upon release to 1 bar and the triple point reference pressures for forsterite and silica. We calculate the corresponding impact velocities by assuming that the projectile and target are both the same material and using impedance matching. The difference between forsterite, one of the most refractory silicate phases, and $\alpha$-quartz, a comparatively less refractory phase, are clearly seen by the large differences in the shock pressures and impact velocities required to initiate melting and vaporization.

Because of the topology of the vapor curve in pressure-entropy space (Fig.~\ref{fig:ent_schem}), the onset of vaporization occurs at a lower specific entropy when decompressing to a lower pressure. 
Depending on the curvature of the liquid-vapor dome, lowering the ambient pressure can dramatically lower the specific entropy at incipient vaporization. In the cases of forsterite and silica, releasing to the triple point instead of 1 bar reduces the incipient vaporization impact velocity by approximately 2 km/s. 
Overall, impact-induced phase changes require lower pressures for the onset of melting and vaporization for collisions within the pressure of the solar nebula, which is near the triple points of silicates, compared to a 1 bar reference pressure. 

{At larger entropies, the effect of the final pressure on the final vapor fraction can reverse because of the asymmetric topology of the vapor dome. The impact criteria for 50\% vaporization increases at the triple point pressure compared to 1 bar, approximately 4 km/s for forsterite, and 3 km/s for silica. We note that the shock pressures required for 50\% vaporization of forsterite upon release to the triple point fall beyond the range of validity of our Hugoniot entropy calculation. The entropies are extrapolated for shock pressures beyond 1 TPa.}

The initial temperature is an important factor. For the warmer initial conditions, forsterite melt{s} and vaporize{s} at much lower shock pressures upon release compared to an initial temperature of 298~K. Warm forsterite (1200 K) can reduce the impact velocity required to begin vaporizing by about {1.5} km/s. It is vital not to neglect the thermal state of impacting materials especially when considering problems in the early solar system.

\begin{table*}
\centering
 \small
 \begin{tabular*}{\textwidth}{@{\extracolsep{\fill}} l l l  l l l} 
 Initial Temp. (K) & Release $P$ (Pa)& Variables &Complete Melt&Incipient Vap.& 50\% Vap.\\ [0.5ex] 
 \hline  

\multicolumn{6}{@{\extracolsep{\fill}} l}{Experimental forsterite (Mg$_2$SiO$_4$), this work} \\ [0.5ex] 
 \hline  
298  & $10^5$ & $P$ (GPa)&  189($\pm$44) & 270($\pm$73) & 856($\pm$189)\\ 
 & & $V_i$ (km/s)  & 9.8($\pm$1.4) & 12.2($\pm$2.0) & 24.3($\pm$3.2)\\ 
 & &$S $ (kJ/K/kg) & 3.474($\pm$.197) & 4.270($\pm$.279)&  6.635\\

298  & $5.2$ & $P$ & 189($\pm$44) &  189($\pm$44) & 1126\\ 
 & & $V_i$  & 9.8($\pm$1.4) & 9.8($\pm$1.4) & 28.6\\ 
 & &$S $ & 3.474($\pm$.197) &3.474($\pm$.197)&  7.616\\

1200 & $10^5$ &  $P$ & 142& 220 & 791\\ 
 & & $V_i$ & 8.2& 10.9 & 23.4\\ 
 & &$S $ & 3.474($\pm$.197) & 4.270($\pm$.279)& 6.635\\

1200  & $5.2$ & $P$  &  142& 142& 1397\\ 
 & & $V_i$  & 8.2 & 8.2 & 32.4\\
 & &$S $ & 3.474($\pm$.197) & 3.474($\pm$.197) & 7.616\\[0.5ex] 

\multicolumn{6}{@{\extracolsep{\fill}} l}{Experimental $\alpha$-quartz (SiO$_2$), revised from \citet{kraus2012shock}} \\ [0.5ex] 
  \hline
298  & $10^5$ & $P$ &  57.6($\pm$2.34) & 90.9($\pm$1.98) & 260($\pm$5.27)\\ 
 & & $V_i$  & 5.89($\pm$.31) &  7.83($\pm$.29) & 14.27($\pm$.39)\\ 
 & &$S $ & 2.950 & 3.560& 5.394\\

298 & $2.68$ & $P$   &  57.6($\pm$2.34) & 57.6($\pm$2.34) & 371($\pm$8.49)\\ 
 & & $V_i$  & 5.89($\pm$.31) & 5.89($\pm$.31) & 17.43($\pm$.45)\\
 & &$S $ & 2.950 & 2.950& 6.100\\

  \hline
\end{tabular*}
\caption{  Critical pressures, impact velocities and specific entropies required  for complete melt, incipient vaporization, and 50\% vapor for two ambient pressures. 1 bar forsterite incipient vaporization entropy is estimated based on {evaporation experiments} \citep{nagahara1994evaporation}{ and the new ANEOS boiling point.} The 50\% vaporization entropies for forsterite are derived from the new ANEOS model vapor curve. All other values are experimentally constrained; see text for details.}
\label{OnsetTable}
\end{table*}

\section{Implications for Planet Formation}

Melting and vaporization can have significant effects on the thermal and chemical evolution of planetesimals and planetary bodies. However, the degree to which silicates are melted and vaporized during planet formation are unknown. Here, we assess the importance of vaporizing collisions that occur during the growth of terrestrial planets. 

Considering the critical impact conditions in Table~\ref{OnsetTable}, {cold} silica and {warm} forsterite begin to vaporize when impact velocities exceed about {6} and {8}~km~s$^{-1}$ respectively. These velocities correspond to the escape velocities of approximately Mars-mass{ and larger} planetary embryos. Thus, essentially all collisions onto warm, differentiated planetary embryos are in {or approaching }the vaporization regime. However, during terrestrial planet formation, most of the mass that impacts onto the largest bodies is accreted because of their deep gravitational potential wells \citep{leinhardt2012,asphaug2010similar}. {Giant impacts involving planetary embryos are extremely high energy events that cause substantial vaporization}. In these giant impacts, the vaporized mantles generate transient silicate vapor atmospheres \citep{lock2017structure,Carter20}. In giant impacts, variable amounts of vaporizing material escape the growing body as ejecta.
In contrast, on smaller bodies, the velocities required for vaporization fall in the erosive or catastrophic disruption regimes \citep{leinhardt2012,carter2019high}. Next, we examine a range of collisions that occur during the growth of the rocky planets in our solar system, starting with a population of small planetesimals.

\subsection{Vaporizing Collisions during Planet Formation}

To investigate the importance of melting and vaporization during accretion, $N$-body planet formation simulations that track collisional fragmentation and re-accretion \citep{carter2015compositional} were post-processed to extract the impact conditions and outcomes.

Here we consider a high resolution, dynamically excited $N$-body simulation based on \citet{carter2015compositional}. This new simulation is a high resolution version of their simulation 27, which includes migration of Jupiter and aerodynamic drag from the nebula. The accretion simulation began with 100\,000 particles distributed between 0.5 and 3\,au, with the smallest bodies having masses of $\sim6 \times 10^{-5}$\,M$_\oplus$ (corresponding to radii of $\sim$200\,km). The mass of collision fragments smaller than this mass limit were placed into an `unresolved debris' annulus corresponding to the location of the collision. Resolved planetesimals and embryos reaccreted this debris as they passed through these annuli, thus recycling the small ejecta \citep{leinhardtandrichardson2005,leinhardt2015,carter2015compositional}. These simulations used particle radius inflation to reduce the computation time, so the impact velocities were first corrected for the additional acceleration associated with infall from the inflated radius to a radius calculated using an assumed density of 3 g\,cm$^{-3}$ (the radius given above is the uninflated value). This correction is only significant for low velocity impacts, and so has a negligible effect on our results.

\begin{figure*}

\includegraphics[width=\textwidth]{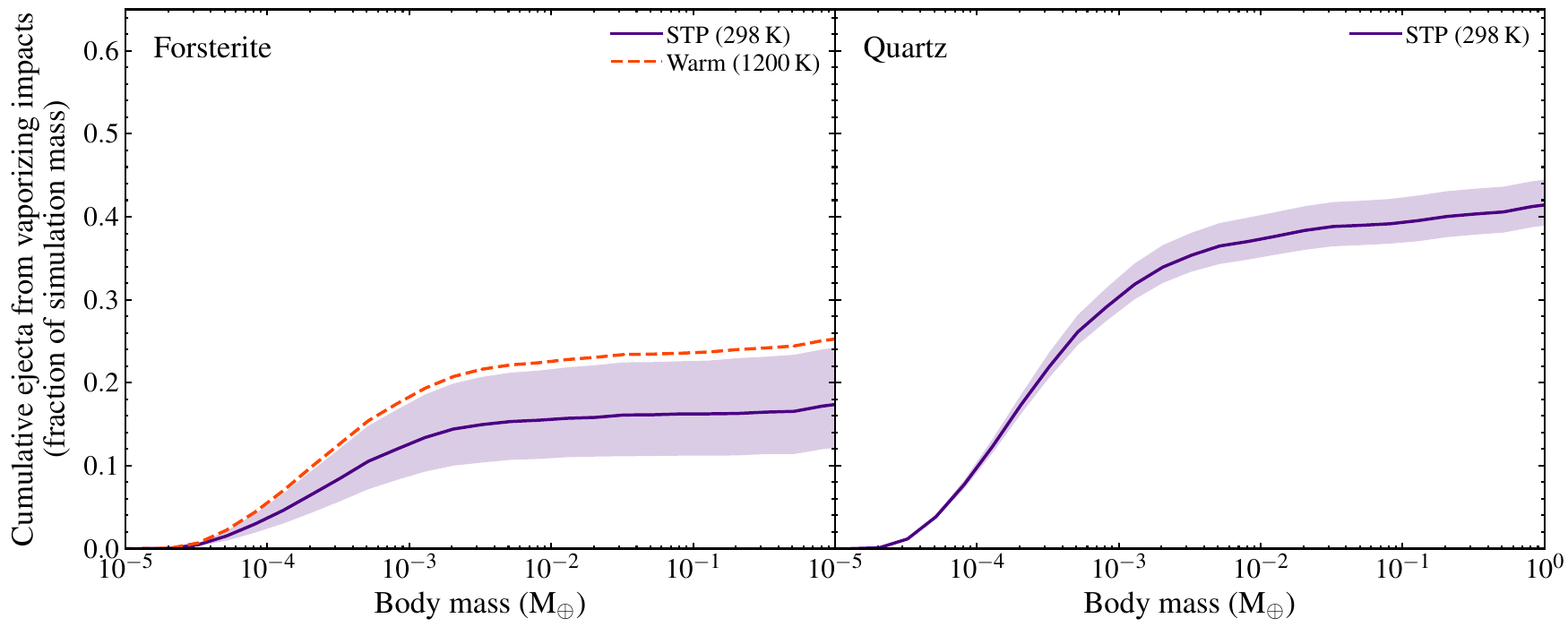}

\caption{Cumulative fraction of total simulation mass ejected in collisions that exceed the critical impact velocities for the onset of vaporization of forsterite (left) and $\alpha$-quartz (right), upon release to the triple point, against the mass of the target body. Purple curves indicate initial conditions at STP, while {the} dashed orange curve indicate{s} {a} warm initial temperature. Warm initial conditions are used to show the dependence on temperature for the amount of melt and vapor that is induced for impacts of the same velocity. A significant fraction of mass is melted and/or vaporized, whether the material is forsterite or $\alpha$-quartz. Up to {20-40}\% of system mass can be processed through impact induced melting and vaporization, and the majority of such collisions occur between planetesimals. Warm initial conditions dramatically increase the amount of mass that is partially vaporized. The shaded regions indicate the range of curves calculated by adding/subtracting the corresponding uncertainties to the critical values given in Table \ref{OnsetTable}. 
This figure uses the impact data from the new, high resolution version of simulation 27 from \citet{carter2015compositional}.}
\label{fig:ejecta}
\end{figure*}

Using the critical impact velocities calculated above for triple point release pressures, we selected all of the collisions in the simulation that exceed the requirements for phase changes upon release to the triple point pressure. We calculated the mass of unresolved `debris' ejected during collisions that exceeded the critical impact velocities. The mass ejected in impacts that cause vaporization is shown as a cumulative sum versus target mass for this example simulation of dynamically excited planet formation in Figure \ref{fig:ejecta}. The steepness of the curves for target bodies between $10^{-4}$ to $10^{-3}$ M$_{\oplus}$ indicates that most of {collisions that exceed the threshold for melting and vaporization occur between planetesimals.}

Figure \ref{fig:ejecta} shows that the effects of warmer initial conditions of impacts are dramatic. Collisions between warm, differentiated planetesimals (dashed orange line) substantially increase{s} the amount of mass processed through melting and vaporization compared to cold planetesimals (purple lines). 
However, the parameters for cold (298 K) forsterite provide a conservative estimate for the mass of material that is processed through vaporizing collisions. Less refractory phases, such as $\alpha$-quartz, melt and vaporize upon release at much lower impact velocities. Warm initial conditions only reduce the required impact velocity further.

The limitations of calculating melting and vaporization in $N$-body simulations need to be addressed. Even in the high resolution $N$-body simulations used in this work, there is not enough information to determine the exact fraction of solid remnant in unresolved or resolved mass. 
Furthermore, in the calculation of ejecta, the thermal state is not taken into account nor is it evolved over time. There is also no differentiating between mass that has been processed multiple times from mass being processed for the first time. Subsequent re-processing of the same mass inflates the total mass that is summed in Figure \ref{fig:ejecta}. Lastly, resolving more mass and smaller planetesimals would allow for more accurate tracking of processed material as compared to large bins of unresolved mass. Resolution is currently computationally limited.

Overall, we expect that up to {20-40}\% of total inner solar system mass can be involved in melting and vaporizing collisions during dynamically excited planet formation such as a Grand Tack scenario. {At present, the cumulative effects of many vaporizing collisions between planetesimals are not well understood. In a Grand Tack scenario, some of the vaporizing collisions would occur within the nebular gas and some after dispersal of the gas. Recent studies found that the outcomes of vaporizing collisions with and without nebular gas will be different in terms of dispersal of material and mass of the largest remnant }\citep{davies2019impact, stewart2019impact, carter2019collapsing}. {In current $N$-body simulations the presence of nebular gas imparts a drag onto the particles, but the differences in collision outcomes are not yet taken into account. In general, collisions between small bodies that produce silicate vapor will also exceed the criteria for catastrophic disruption, where the largest fragment is less than half the total mass.  As a result, the material from the colliding planetesimals is dispersed as many smaller fragments. Note that the available criteria for catastrophic disruption have been calculated for collisions in the absence of a surrounding gas} \citep{leinhardt2012}.

{Our results also have implications for observations of planet formation in exoplanetary systems. Several works }\citep[e.g.][]{Meng14,Su19}{ have suggested that variable emission seen from dust in protoplanetary disks (extreme debris disks) is the result of condensed vapor produced by high energy collisions between growing planets. The critical impact velocities for vaporizing collisions during planet growth that we have calculated here provide key information for interpreting such observations. Our results will allow constraints to be placed on the impact velocities and types of impacts occurring in time-variable disks.}

{While most of the material involved in vaporizing collisions will ultimately be accreted onto the final planets, partial vaporization offers the opportunity for some chemical and isotopic fractionation. Although our calculations of total mass involved in vaporizing collisions is not the same as the amount of vapor produced,  our work demonstrates that major silicate phases can be vaporized. As a result, more volatile components will also be vaporized during these energetic collisions. Even in cases of small vapor fractions, the volume increase is many orders of magnitude, so the onset of vaporization is key for changing the dynamics of the ejecta. The total mass of vapor produced over time, and the composition of that vapor, requires additional information about the distribution of shock pressures within the colliding bodies and the compositions of the bodies. Whether or not a bulk chemical and isotopic fractionation would be imparted onto the final planets remains to be determined, as it depends on how efficiently the shock-processed planetesimal material is accreted. This work provides the motivation to pursue these more detailed investigations. }

\section{Conclusions}
New experimental capabilities in shock physics provide thermodynamic data that is essential for understanding planetary materials and planet formation processes. Precise P-V-T Hugoniot measurements provide powerful constraints on the equation of state of materials, which enable a more robust understanding of material properties during planet formation. In this work, we have presented key thermodynamic data on forsterite and used our constraints on the EOS to evaluate the conditions required for shock-induced melting and vaporization during planetary collisions. Previously developed ANEOS forsterite model equations of state currently in use in hydrocode simulations of collisions generally under predict production of entropy on the forsterite principal Hugoniot over the shock pressures encountered during planet formation. A revised ANEOS model for forsterite has been developed with our new data.

The calculations presented here offer constraints on the amount of melting and vaporization during planet formation. Cycling of mass through ejecta and re-accretion is a common process throughout planet formation, and impact velocities above the vaporization criteria are common as well. We have shown that the cycled mass can be a significant fraction of the total mass of the final terrestrial planets. Therefore, processing of planetesimals via impact induced melting and vaporization must be considered prevalent during terrestrial planet formation. Further investigations are required to determine the full effects of vaporizing collisions.

To make more precise calculations of partially vaporized materials, we need experiments that measure the liquid-vapor boundary for forsterite. The vapor curve has been measured for silica and iron using shock-and-release techniques \citep{kraus2012shock,kraus2015impact}, and similar experiments on forsterite are in progress. {The addition of temperature dependence on the Gr\"uneisen parameter model would further refine this work as well.} Future coupling of equations of state data to $N$-body simulations would allow for the tracking of specific entropy and energy throughout planet formation.

\acknowledgments
This work was conducted under the Z Fundamental Science Program. The authors thank the support from DOE-NNSA grant DE-NA0002937, NASA grants NNX15AH54G and NNX16AP35H, UC Office of the President grant LFR-17-449059, and DOE-NNSA grant DE-NA0003842.  Sandia National Laboratories is a multimission laboratory managed and operated by National Technology and Engineering Solutions of Sandia, LLC., a wholly owned subsidiary of Honeywell International, Inc., for the U.S. Department of Energy's National Nuclear Security Administration under contract DE-NA0003525.  This paper describes objective technical results and analysis.  Any subjective views or opinions that might be expressed in the paper do not necessarily represent the views of the U.S. Department of Energy or the United States Government. This work was performed under the auspices of the U.S. Department of Energy by Lawrence Livermore National Laboratory under Contract DE-AC52-07NA27344. The revised ANEOS parameter set for forsterite, modifications to the ANEOS code, and SESAME and GADGET format EOS tables are available at {https://github.com/isale-code/M-ANEOS and  https://github.com/ststewart/aneos-forsterite-2019 DOI: 10.5281/zenodo.3478631}. All scripts used for calculations in this work are available at https://github.com/edavi006/Forsterite$\_$entropy$\_$2019 {DOI: 10.5281/zenodo.3610687.}

%Supporting information for Silicate Melting and vaporization during rocky planet formation

%make all the sections labelled with an S for supp
\renewcommand{\thepage}{S\arabic{page}}  
\renewcommand{\thesection}{S\arabic{section}}   
\renewcommand{\thetable}{S\arabic{table}}   
\renewcommand{\thefigure}{S\arabic{figure}}
\renewcommand{\theequation}{S\arabic{equation}}

%reset page count and section count
\setcounter{page}{1}
\setcounter{section}{0}
\setcounter{figure}{0}
\setcounter{table}{0}
\setcounter{equation}{0}

 \cleardoublepage
\newpage
\appendix{Supporting Information for Silicate Melting and Vaporization during Rocky Planet Formation}
% \author{E. J. Davies\affil{1}, P. J. Carter\affil{1}, S. Root\affil{2}, R. G. Kraus\affil{3}, D. K. Spaulding\affil{1}, S. T. Stewart\affil{1}, S. B. Jacobsen\affil{4}}

%\affiliation{1}{Department of Earth and Planetary Sciences, U. California, Davis, CA, USA}
%\affiliation{2}{Sandia National Laboratories, Albuquerque, NM, USA}
%affiliation{3}{Lawrence Livermore National Laboratory, Livermore, CA, USA}
%\affiliation{4}{Department of Earth and Planetary Science, Harvard University, Cambridge, MA, USA}

%\email{Erik Davies}{ejdavies@ucdavis.edu}

\begin{enumerate}
\item Supporting Text S1-S4.
\item Figure S1-S3.
\item Table S1.
\end{enumerate}

\section{Revised Critical Shock Pressures for Shock-Induced Phase Changes in Quartz}\label{Sup:quartzSection}

The vaporization criteria for quartz (SiO$_2$) are based on the experiments and absolute entropy calculations in \cite{kraus2012shock}. We used the revised M-ANEOS vapor curve from \cite{kraus2012shock}. This is a conservative value for the onset of vaporization upon release to the triple point because we have neglected the enthalpy of melting crystobalite, which leads to a reduction in entropy of about 74~J/K/kg \citep{richet1982thermodynamic} for the onset of sublimation. The triple point pressure of about 2.6~Pa for SiO$_2$ was experimentally determined by \cite{mysen1988condensation}. The fitted equation for entropy on the $\alpha$-quartz Hugoniot from \cite{kraus2012shock} (Eq.~7) was valid from 110 to 800 GPa. Here, we extend the fit to lower pressures, into the solid region along the Hugoniot using the stishovite equation of state model in \cite{kraus2012shock}, the results is shown in Table \ref{tab:equations}. 

In deriving the associated impact velocities in Table~\ref{OnsetTable}, we used the Hugoniot from \cite{knudson2013adiabatic}. Note that the critical pressures in Table~6 of \cite{kraus2012shock} are incorrect. After having checked all the entries for the shock pressures to vaporize quartz presented in Table 6 of \cite{kraus2012shock}, we found an error in the calculation of the shock pressure to achieve incipient vaporization upon release to 2 and 10$^5$ Pa due to a typographical error in the determination of the entropy along the quartz Hugoniot (using an anharmonicity of 4.6 instead of 3.1).  This typographical error does not impact the entropy calculation in the fluid or the equations presented in text as the entropy along the fused quartz Hugoniot was used to determine the reference entropy.

\section{Update to Forsterite Temperatures on Z}\label{ForsteriteDatUp}

During analysis of data in \cite{root2018forsterite}, the streaked visible spectroscopy (SVS) records from Z3033 North 7 and South 7 were inadvertantly swapped. Furthermore, South 5 was reported as having a shock temperature when the temperature actually came from South 7 from the same experiment. A reanalysis of the data with the correct orientations of the records yields the following results for the shocked state: for Z3033 North 7, $P$ = 496.8$\pm$4.8 GPa, $T$ = 16104$\pm$1653 K, and $u_s$ = 17.47$\pm$.21 km/s; and for Z3033 South 7, $P$ = 564.6$\pm$5.3 GPa, $T$ = 20229$\pm$1969 K, and $u_s$ = 18.33$\pm$.16 km/s. South 5 is the same as reported in \citet{root2018forsterite}, except with no associated temperature. See \citet{root2018forsterite} for more information on the analysis process.

\section{Gr{\"u}neisen Parameter Formulations}\label{Sup:gammasection}

The formulations of $\gamma$ used in this work include an exponential, $\gamma(\rho) =\gamma_0 \left(\frac{\rho_0}{\rho}\right)^q$, as in \citet{thomas2013direct}, a linear model, $\gamma(\rho) = \gamma_0 + \gamma' \left( \frac{\rho_0}{\rho} - 1\right)$, as in \cite{de2008thermodynamics} where $\gamma'$ takes the place of $q$, a linear model with a high initial $\gamma$ ($>$ 2), and a Gaussian. The first two formulations have been used to model $\gamma$ to densities of twice the initial density of the liquid. Under the assumption of $\gamma$'s dependence on density, $\gamma$ across all of the decompression isentropes must overlap in density space. We found that the formulations of $\gamma$ from \citet{thomas2013direct} and \citet{de2008thermodynamics} do not meet this criteria for the release isentropes calculated here. The density range of this data set is far beyond the densities considered in those studies.

The linear formulation with a high initial $\gamma$ does better, however $\gamma$ often becomes negative at the higher densities of the shocked state. We found that a Gaussian of the form $\gamma=0.5+1.2(0.5)e^{(q(\rho-5000(500))^2)}$ does best to fit the high density data while keeping $\gamma$ positive. Parameters on the Gaussian are allowed to vary randomly about a normal distribution over the parameter space until convergence is achieved. There is also a strong dependence on the calculated volume of the release state. Regardless of formulation, $\gamma$ at the final release state is similar between all formulations however, uncertainties are larger for formulations that do not converge well. In all considered formulations, the resulting isentropic path is similar, so the change in energy between different formulations is relatively small, therefore $\gamma$ is similar. 

\begin{figure*}[hb]
\centering
\includegraphics[width=5in]{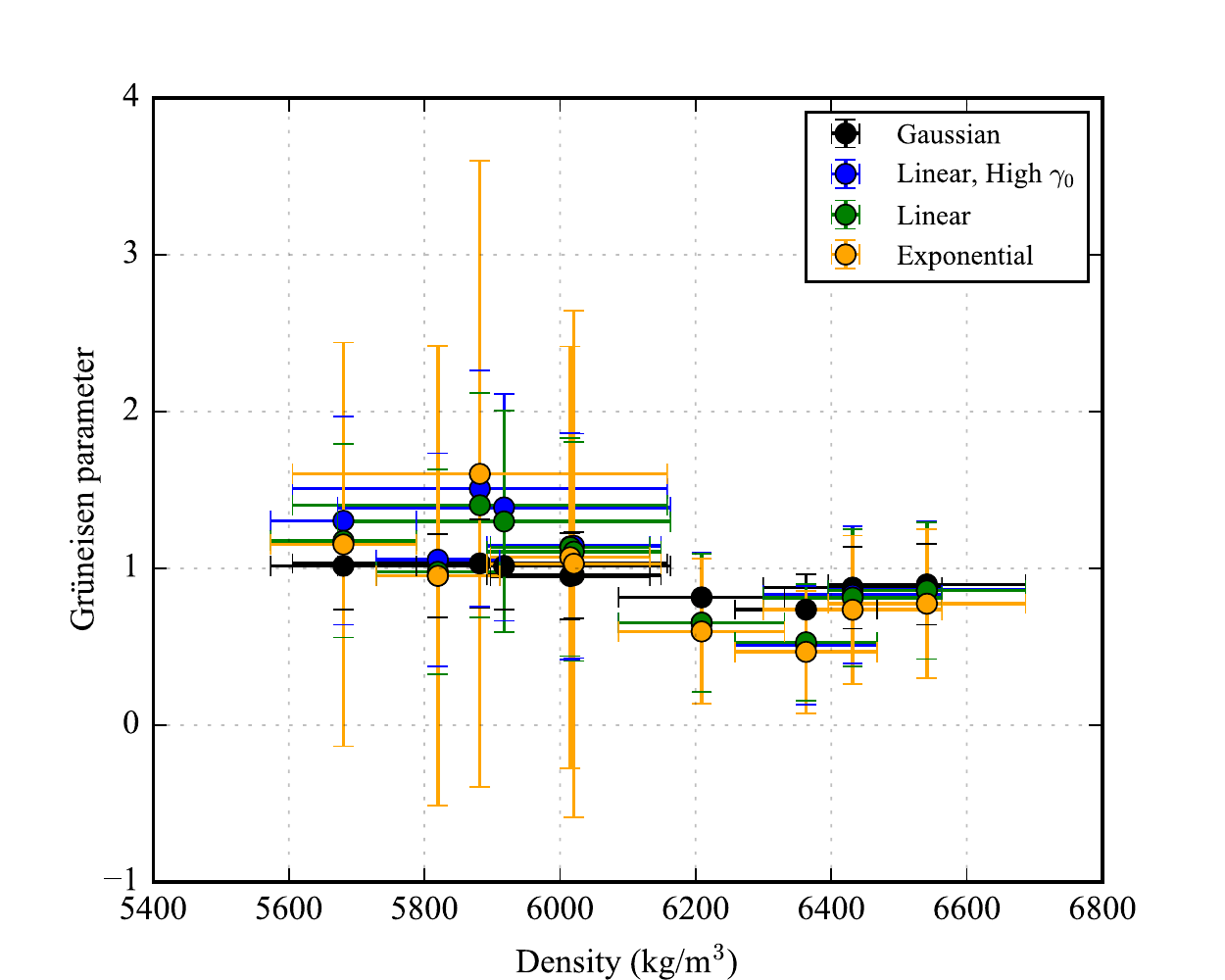}
\includegraphics[width=5in]{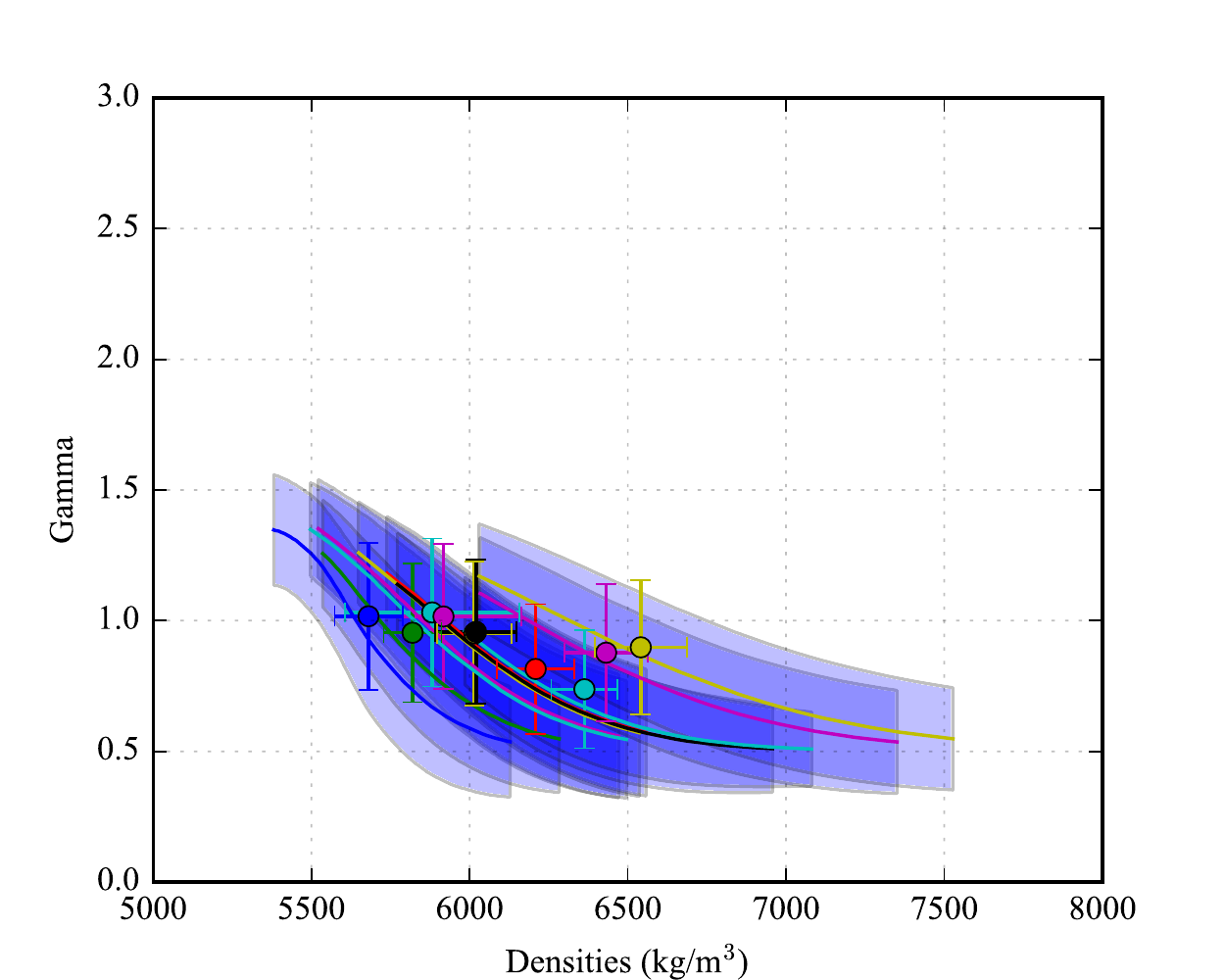}

\caption{ (Top) Calculated $\gamma$ across all considered formulations. Uncertainty for the exponential, and two linear models are large, but points still cluster together. The preferred Gaussian model is within uncertainty of all three other models. (Bottom) $\gamma$'s dependence on density for each release isentrope. Paths largely overlap within error and final release states (points) cluster together at similar densities. }
\label{Sfig:Gammas_paths}
\end{figure*}

Figure \ref{Sfig:Gammas_paths} shows results of all considered formulations for $\gamma$ in release isentrope calculations and continuous calculated $\gamma$'s with the Gaussian formulation during release.

To fit the available and measured values of $\gamma$, the main text describes a function of the form $\gamma(\rho) = \gamma_{\infty} + (A - \gamma_{\infty}) \left( \frac{\rho_0}{\rho} \right)^{B} + C e^{-(\rho - D)^2/E^2}$. We fit through the available data and their uncertainties to get best fit parameters. Assuming the same function topology, we assert a constant percentile error of gamma at each density. The percentile error is chosen by generating statistical data clouds using the uncertainty about each data point and making sure the chosen percentile error about the fit contains approximately 66\% of the total points in the data clouds. A percentile error of 32\% satisfies these conditions.

\section{Planet Formation Eccentricity}

\begin{figure*}
        \centering
        \includegraphics[width=\textwidth]{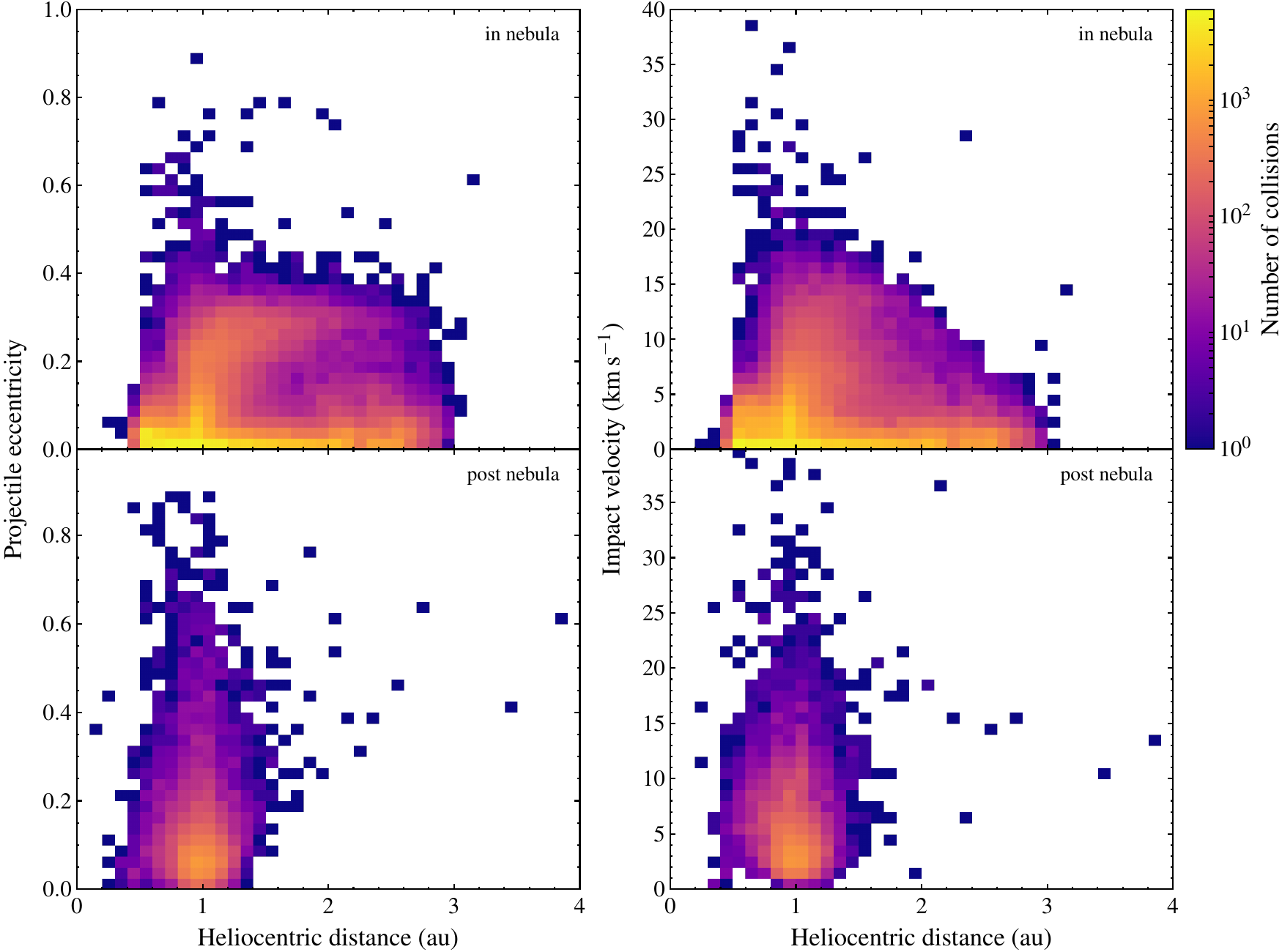}
%%
% Equivalent total cumulative debris mass for calm sim:
% During: 0.19 Earth masses; After: 0.42 Earth masses
%%

        \caption{Eccentricity of projectile (left) in impacts and impact velocity (right) histograms as a function of heliocentric distance, while the nebula is present (top), and after the nebula disperses (bottom). While embedded in the nebula, most of the impacts have low eccentricity due to damping by the gas. However, there is a large increase in eccentricity once Jupiter migrates during the Grand Tack. Once the nebula disperses, there is no gas to damp the planetesimals, so eccentricities remain high relative to the total number of collisions.
        In summary, eccentricities remain high, and the number of collisions remains large even after the nebula has dispersed.
        This figure uses the impact data from the new, high resolution version of simulation 27 from \citet{carter2015compositional}.}
        \label{sup:fig:ecentricity}
    \end{figure*}

\begin{figure*}
        \centering
        \includegraphics{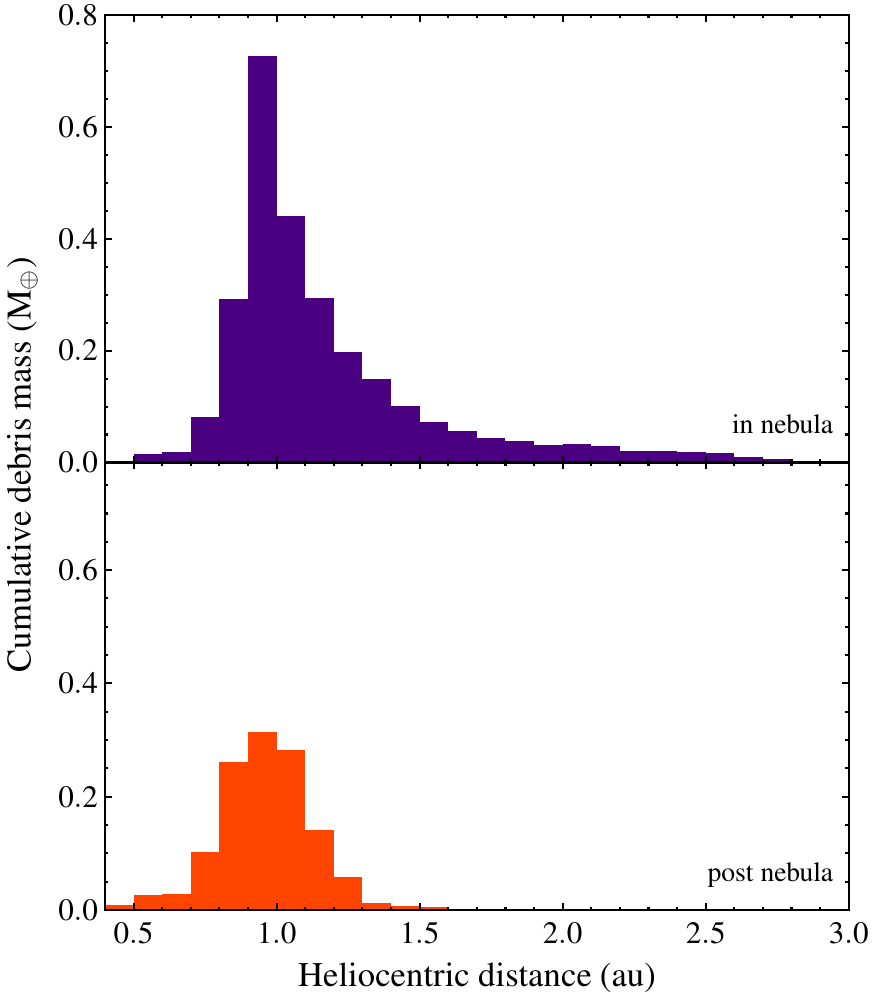}
        \caption{Cumulative mass of ejecta generated from impacts over the duration of the $N$-body simulation as a function of heliocentric distance. The top panel shows the binned debris mass produced while the nebula is present, the bottom panel shows the debris mass produced after the nebula has dispersed. There are many impacts that generate debris both during and after the nebula phase. This figure uses the impact data from the new, high resolution version of simulation 27 from \citet{carter2015compositional}.}
        \label{sup:fig:semiMass}
    \end{figure*}

In the nebula, aerodynamic drag dampens the eccentricities of planetesimals, reducing the relative velocity between them (Fig. \ref{sup:fig:ecentricity}). Initially, gas drag keeps the eccentricities low, and most collisions occur between low eccentricity planetesimals. During the Grand Tack, there is an initial spike in eccentricities and impact velocities as giant planet migration initiates. After the migration, the planetesimals in the inner solar system remain at high eccentricities, leading to elevated impact velocities long after the migration has ended. Other scenarios of planet formation such as an early Nice migration, or sweeping resonances from the giant planets, can also produce the orbital eccentricities required to instigate collisions with sufficient velocities to enter the melting and vaporization regimes. Figure \ref{sup:fig:semiMass} shows the cumulative debris mass generated during impacts with respect to heliocentric distance. The total mass ejected in the early solar system is counted in Earth masses. The outcomes of these early collisions are dependent on the pressures in the nebula that the impact pressures release to, for which the triple point pressure is a proxy. %These collisional outcomes in the nebula are discussed in Stewart et al. (submitted). 

Once the nebula disperses, dynamic friction transfers angular momentum from larger bodies to smaller bodies, increasing their eccentricity. Projectile eccentricity during collisions after the nebula disperses is shown in the second panel of Figure \ref{sup:fig:ecentricity}. It is clear that there is a large variety of eccentricities, and therefore velocities,  around 1 au post-nebula. From $N$-body simulations, it is shown in Figure \ref{sup:fig:semiMass} that the cumulative debris mass ejected in collisions after nebular dispersal is also large. Because eccentricities are large post-nebula, vaporizing collisions still occur and because the nebula has dispersed, the ambient pressure in the inner solar system has decreased as well.

\section{Supplementary Table}

Additional table to support the text.

\begin{center}
\begin{table*}
\centering
 \small
\begin{tabular*}{\textwidth}{l @{\extracolsep{\fill}}  l }
 \hline

       Forsterite $u_s(u_p)$ & $\left[ \begin{array}{cccc}  1.831 &  -6.523E{-01}  &7.319E{-02} & -2.609E{-03}\\ -6.523E{-01}  & 2.366E{-01}  &-2.685E{-02}&  9.671E{-04}\\7.319E{-02} & -2.685E{-02}  & 3.084E{-03}  & -1.122E{-04}\\-2.609E{-03}& 9.671E{-04} &  -1.122E{-04} &  4.125E{-06} \end{array}\right]$\\ 

    Forsterite $S(P)$ & $\left[ \begin{array}{cccc}  
       1.970E{07} &  5.855E{04}  &-1.097E{01}&  -8.471E{04}\\
       5.855E{04}  & 1.782E{02}&  -3.494E{-02} & -1.724E{02}\\
      -1.097E{01}  &-3.494E{-02}&   7.447E{-06}  & 1.931E{-03}\\
      -8.471E{04} & -1.724E{02}&   1.931E{-03}  & 1.911E{03} \end{array}\right]$\\ 
   
   Forsterite $T(u_s)$ & $\left[ \begin{array}{ccc} 1.656E05 & -2.027$E$04 &5.987$E$02 \\ -2.027E04 & 2.530$E$03  &-7.263E01 \\ 5.987E02  & -7.263E01 & 2.357\end{array}\right]$ \\ 
   
 \hline  
       Quartz $u_s(u_p)$ & $\left[ \begin{array}{cccc}  2.097E{-02} &  -6.159E{-03}  &5.566E{-04} & -1.572E{-05}\\ -6.159E{-03}  & 1.877E{-03}  &-1.742E{-04}&  5.017E{-06}\\5.566E{-04} & -1.742E{-04}  & 1.65E{-05}  & -4.834E{-07}\\-1.572E{-05}& 5.017E{-06} &  -4.834E{-07} &  1.441E{-08} \end{array}\right]$\\ 
 \hline
        Fused Silica $u_s(u_p)$ & $\left[ \begin{array}{cccc}  4.857E{-02} &  -1.343E{-02}  &1.17E{-03} & -3.2366E{-05}\\ -1.343E{-02}  & 3.84E{-03}  &-3.4252E{-04}&  9.6408E{-06}\\1.17E{-03} & -3.4252E{-04}  & 3.1126E{-05}  & -8.886E{-07}\\-3.2366E{-05}& 9.6408E{-06} &  -8.886E{-07} &  2.5662E{-08} \end{array}\right]$\\ 
 \hline
\end{tabular*}
\caption{Covarience matrices for equations in Table \ref{tab:equations}. The parameters in each equation map to the covariance matrices from left to right.}
\label{tab:fo_fit}
\end{table*}
\end{center}


\begin{thebibliography}{76}
\providecommand{\natexlab}[1]{#1}
\expandafter\ifx\csname urlstyle\endcsname\relax
  \providecommand{\doi}[1]{doi:\discretionary{}{}{}#1}\else
  \providecommand{\doi}{doi:\discretionary{}{}{}\begingroup
  \urlstyle{rm}\Url}\fi

\bibitem[{\textit{Ahrens and O'Keefe}(1972)}]{ahrens1972shock}
Ahrens, T.~J., and J.~D. O'Keefe (1972), Shock melting and vaporization of
  lunar rocks and minerals, \textit{The Moon}, \textit{4}(1-2), 214--249.

\bibitem[{\textit{Asimow}(2018)}]{asimow2018melts}
Asimow, P.~D. (2018), Melts under extreme conditions from shock experiments, in
  \textit{Magmas Under Pressure}, pp. 387--418, Elsevier.

\bibitem[{\textit{Asphaug}(2010)}]{asphaug2010similar}
Asphaug, E. (2010), Similar-sized collisions and the diversity of planets,
  \textit{Chemie der Erde-Geochemistry}, \textit{70}(3), 199--219.

\bibitem[{\textit{Belonoshko et~al.}(2005)\textit{Belonoshko, Skorodumova,
  Rosengren, Ahuja, Johansson, Burakovsky, and Preston}}]{belonoshko2005high}
Belonoshko, A.~B., N.~Skorodumova, A.~Rosengren, R.~Ahuja, B.~Johansson,
  L.~Burakovsky, and D.~Preston (2005), High-pressure melting of {MgSiO} 3,
  \textit{Physical review letters}, \textit{94}(19), 195,701.

\bibitem[{\textit{Boehly et~al.}(1994)\textit{Boehly, Craxton, Hinterman,
  Jaanimagi, Kelly, Kessler, Kremens, Kumpan, Letzring, McCrory
  et~al.}}]{boehly1994upgrade}
Boehly, T., R.~Craxton, T.~Hinterman, P.~Jaanimagi, J.~Kelly, T.~Kessler,
  R.~Kremens, S.~Kumpan, S.~Letzring, R.~McCrory, et~al. (1994), The upgrade to
  the omega laser system, \textit{Fusion technology}, \textit{26}(3P2),
  722--729.

\bibitem[{\textit{Bonsor et~al.}(2015)\textit{Bonsor, Leinhardt, Carter,
  Elliott, Walter, and Stewart}}]{bonsor2015collisional}
Bonsor, A., Z.~M. Leinhardt, P.~J. Carter, T.~Elliott, M.~J. Walter, and S.~T.
  Stewart (2015), A collisional origin to {Earth}$\lq$s
  non-chondritic composition?, \textit{Icarus}, \textit{247}, 291--300.

\bibitem[{\textit{Bouhifd et~al.}(1996)\textit{Bouhifd, Andrault, Fiquet, and
  Richet}}]{bouhifd1996thermal}
Bouhifd, M.~A., D.~Andrault, G.~Fiquet, and P.~Richet (1996), Thermal expansion
  of forsterite up to the melting point, \textit{Geophysical Research Letters},
  \textit{23}(10), 1143--1146.

\bibitem[{\textit{Burakovsky and Preston}(2004)}]{burakovsky2004analytic}
Burakovsky, L., and D.~L. Preston (2004), Analytic model of the {G}r{\"u}neisen
  parameter all densities, \textit{Journal of Physics and Chemistry of Solids},
  \textit{65}(8-9), 1581--1587.

\bibitem[{\textit{Canup et~al.}(2013)\textit{Canup, Barr, and
  Crawford}}]{canup2013lunar}
Canup, R., A.~Barr, and D.~Crawford (2013), Lunar-forming impacts:
  {h}igh-resolution {SPH} and {AMR-CTH} simulations, \textit{Icarus},
  \textit{222}(1), 200--219.

\bibitem[{\textit{Canup}(2012)}]{canup2012forming}
Canup, R.~M. (2012), Forming a moon with an {E}arth-like composition via a
  giant impact, \textit{Science}, \textit{338}(6110), 1052--1055.

\bibitem[{\textit{Carter et~al.}(2019{\natexlab{a}})\textit{Carter, Davies,
  Lock, and Stewart}}]{carter2019high}
Carter, P., E.~Davies, S.~Lock, and S.~Stewart (2019{\natexlab{a}}), High
  collision velocities between planetesimals during planet growth and
  migration, in \textit{Lunar and Planetary Science Conference}, vol.~50.

\bibitem[{\textit{Carter et~al.}(2019{\natexlab{b}})\textit{Carter, Davies,
  Lock, and Stewart}}]{carter2019collapsing}
Carter, P., E.~Davies, S.~Lock, and S.~Stewart (2019{\natexlab{b}}), Collapsing
  impact vapor plumes: A new planetesimal formation environment, in
  \textit{Lunar and Planetary Science Conference}, vol.~50.

\bibitem[{\textit{Carter et~al.}(2015)\textit{Carter, Leinhardt, Elliott,
  Walter, and Stewart}}]{carter2015compositional}
Carter, P.~J., Z.~M. Leinhardt, T.~Elliott, M.~J. Walter, and S.~T. Stewart
  (2015), Compositional evolution during rocky protoplanet accretion,
  \textit{The Astrophysical Journal}, \textit{813}(1), 72.

\bibitem[{\textit{Carter et~al.}(2018)\textit{Carter, Leinhardt, Elliott,
  Stewart, and Walter}}]{carter2018collisional}
Carter, P.~J., Z.~M. Leinhardt, T.~Elliott, S.~T. Stewart, and M.~J. Walter
  (2018), Collisional stripping of planetary crusts, \textit{Earth and
  Planetary Science Letters}, \textit{484}, 276--286.

\bibitem[{\textit{Carter et~al.}(2020)\textit{Carter, Lock, and
  Stewart}}]{Carter20}
Carter, P.~J., S.~J. Lock, and S.~T. Stewart (2020), The energy budgets of
  giant impacts, \textit{Journal of Geophysical Research: Planets},
  \textit{125}(1), e2019JE006,042, \doi{10.1029/2019JE006042}, e2019JE006042
  10.1029/2019JE006042.

\bibitem[{\textit{Celliers et~al.}(2004)\textit{Celliers, Bradley, Collins,
  Hicks, Boehly, and Armstrong}}]{celliers2004line}
Celliers, P., D.~Bradley, G.~Collins, D.~Hicks, T.~Boehly, and W.~Armstrong
  (2004), Line-imaging velocimeter for shock diagnostics at the {OMEGA} laser
  facility, \textit{Review of scientific instruments}, \textit{75}(11),
  4916--4929.

\bibitem[{\textit{Chambers}(2001)}]{chambers2001making}
Chambers, J. (2001), Making more terrestrial planets, \textit{Icarus},
  \textit{152}(2), 205--224.

\bibitem[{\textit{Chambers}(2010)}]{chambers2010terrestrial}
Chambers, J. (2010), \textit{Terrestrial planet formation}, University of
  Arizona Press, Tucson, Arizona, United States (297--317).

\bibitem[{\textit{Chambers}(2013)}]{chambers2013late}
Chambers, J. (2013), Late-stage planetary accretion including hit-and-run
  collisions and fragmentation, \textit{Icarus}, \textit{224}(1), 43--56.

\bibitem[{\textit{Chase et~al.}(1998)\textit{Chase, Davies, Downey, Frurip,
  Mcdonald, and Syverud}}]{chase1998nist}
Chase, M.~W., C.~A. Davies, J.~R. Downey, D.~J. Frurip, R.~A. Mcdonald, and
  A.~N. Syverud (1998), {NIST-JANAF} thermodynamic tables, \textit{J. Phys.
  Chem. Ref. Data Monograph}, \textit{9}.

\bibitem[{\textit{Collins and Melosh}(2014)}]{collins2014improvements}
Collins, G.~S., and H.~J. Melosh (2014), Improvements to {ANEOS} for multiple
  phase transitions, in \textit{Lunar Planet. Sci. Conf}, vol. 2664.

\bibitem[{\textit{{\'C}uk and Stewart}(2012)}]{cuk2012making}
{\'C}uk, M., and S.~T. Stewart (2012), Making the {M}oon from a fast-spinning
  {E}arth: {A} giant impact followed by resonant despinning, \textit{Science},
  \textit{338}(6110), 1047--1052.

\bibitem[{\textit{Davies et~al.}(2019)\textit{Davies, Carter, Duncan, Root,
  Spaulding, Kraus, Stewart, and Jacobsen}}]{davies2019impact}
Davies, E., P.~Carter, M.~Duncan, S.~Root, D.~Spaulding, R.~Kraus, S.~Stewart,
  and S.~Jacobsen (2019), Impact generated vapor plumes after dispersal of the
  solar nebula, in \textit{Lunar and Planetary Science Conference}, vol.~50.

\bibitem[{\textit{de~Koker et~al.}(2008)\textit{de~Koker, Stixrude, and
  Karki}}]{de2008thermodynamics}
de~Koker, N.~P., L.~Stixrude, and B.~B. Karki (2008), Thermodynamics,
  structure, dynamics, and freezing of {Mg$_2$SiO$_4$} liquid at high pressure,
  \textit{Geochimica et Cosmochimica Acta}, \textit{72}(5), 1427--1441.

\bibitem[{\textit{Genda et~al.}(2011)\textit{Genda, Kokubo, and
  Ida}}]{genda2011merging}
Genda, H., E.~Kokubo, and S.~Ida (2011), Merging criteria for giant impacts of
  protoplanets, \textit{The Astrophysical Journal}, \textit{744}(2), 137.

\bibitem[{\textit{Gillet et~al.}(1991)\textit{Gillet, Richet, Guyot, and
  Fiquet}}]{gillet1991high}
Gillet, P., P.~Richet, F.~Guyot, and G.~Fiquet (1991), High-temperature
  thermodynamic properties of forsterite, \textit{Journal of Geophysical
  Research: Solid Earth}, \textit{96}(B7), 11,805--11,816.

\bibitem[{\textit{Izidoro et~al.}(2016)\textit{Izidoro, Raymond, Pierens,
  Morbidelli, Winter, and Nesvorny}}]{izidoro2016asteroid}
Izidoro, A., S.~N. Raymond, A.~Pierens, A.~Morbidelli, O.~C. Winter, and
  D.~Nesvorny (2016), The asteroid belt as a relic from a chaotic early solar
  system, \textit{The Astrophysical Journal}, \textit{833}(1), 40.

\bibitem[{\textit{Jackson and Ahrens}(1979)}]{jackson1979shock}
Jackson, I., and T.~J. Ahrens (1979), Shock wave compression of single-crystal
  forsterite, \textit{Journal of Geophysical Research: Solid Earth},
  \textit{84}(B6), 3039--3048.

\bibitem[{\textit{Knudson and Desjarlais}(2013)}]{knudson2013adiabatic}
Knudson, M., and M.~Desjarlais (2013), Adiabatic release measurements in
  $\alpha$-quartz between 300 and 1200 {GPa}: {C}haracterization of
  $\alpha$-quartz as a shock standard in the multimegabar regime,
  \textit{Physical Review B}, \textit{88}(18), 184,107.

\bibitem[{\textit{Kraus et~al.}(2012)\textit{Kraus, Stewart, Swift, Bolme,
  Smith, Hamel, Hammel, Spaulding, Hicks, Eggert et~al.}}]{kraus2012shock}
Kraus, R., S.~Stewart, D.~Swift, C.~Bolme, R.~Smith, S.~Hamel, B.~Hammel,
  D.~Spaulding, D.~Hicks, J.~Eggert, et~al. (2012), Shock vaporization of
  silica and the thermodynamics of planetary impact events, \textit{Journal of
  Geophysical Research: Planets}, \textit{117}(E9).

\bibitem[{\textit{Kraus et~al.}(2015)\textit{Kraus, Root, Lemke, Stewart,
  Jacobsen, and Mattsson}}]{kraus2015impact}
Kraus, R.~G., S.~Root, R.~W. Lemke, S.~T. Stewart, S.~B. Jacobsen, and T.~R.
  Mattsson (2015), Impact vaporization of planetesimal cores in the late stages
  of planet formation, \textit{Nature Geoscience}, \textit{8}(4), 269.

\bibitem[{\textit{{Leinhardt} and
  {Richardson}}(2005)}]{leinhardtandrichardson2005}
{Leinhardt}, Z.~M., and D.~C. {Richardson} (2005), Planetesimals to
  protoplanets. {I}. {E}ffect of fragmentation on terrestrial planet formation,
  \textit{The Astrophysical Journal}, \textit{625}, 427--440.

\bibitem[{\textit{{Leinhardt} and {Stewart}}(2012)}]{leinhardt2012}
{Leinhardt}, Z.~M., and S.~T. {Stewart} (2012), Collisions between
  gravity-dominated bodies. {I}. {O}utcome regimes and scaling laws,
  \textit{The Astrophysical Journal}, \textit{745}, 79.

\bibitem[{\textit{{Leinhardt} et~al.}(2015)\textit{{Leinhardt}, {Dobinson},
  {Carter}, and {Lines}}}]{leinhardt2015}
{Leinhardt}, Z.~M., J.~{Dobinson}, P.~J. {Carter}, and S.~{Lines} (2015),
  Numerically predicted indirect signatures of terrestrial planet formation,
  \textit{The Astrophysical Journal}, \textit{806}, 23.

\bibitem[{\textit{Lemke et~al.}(2003)\textit{Lemke, Knudson, Hall, Haill,
  Desjarlais, Asay, and Mehlhorn}}]{lemke2003characterization}
Lemke, R., M.~Knudson, C.~Hall, T.~Haill, P.~Desjarlais, J.~Asay, and
  T.~Mehlhorn (2003), Characterization of magnetically accelerated flyer
  plates, \textit{Physics of Plasmas}, \textit{10}(4), 1092--1099.

\bibitem[{\textit{Lemke et~al.}(2005)\textit{Lemke, Knudson, Bliss, Cochrane,
  Davis, Giunta, Harjes, and Slutz}}]{lemke2005magnetically}
Lemke, R., M.~Knudson, D.~Bliss, K.~Cochrane, J.-P. Davis, A.~Giunta,
  H.~Harjes, and S.~Slutz (2005), Magnetically accelerated, ultrahigh velocity
  flyer plates for shock wave experiments, \textit{Journal of Applied Physics},
  \textit{98}(7), 073,530.

\bibitem[{\textit{Lemke et~al.}(2011)\textit{Lemke, Knudson, and
  Davis}}]{lemke2011magnetically}
Lemke, R.~W., M.~D. Knudson, and J.-P. Davis (2011), Magnetically driven
  hyper-velocity launch capability at the {Sandia} {Z} accelerator,
  \textit{International Journal of Impact Engineering}, \textit{38}(6),
  480--485.

\bibitem[{\textit{Lock and Stewart}(2017)}]{lock2017structure}
Lock, S.~J., and S.~T. Stewart (2017), The structure of terrestrial bodies:
  {I}mpact heating, corotation limits, and synestias, \textit{Journal of
  Geophysical Research: Planets}, \textit{122}(5), 950--982.

\bibitem[{\textit{Lodders}(2003)}]{lodders2003solar}
Lodders, K. (2003), Solar system abundances and condensation temperatures of
  the elements, \textit{The Astrophysical Journal}, \textit{591}(2), 1220.

\bibitem[{\textit{Lyzenga and Ahrens}(1980)}]{lyzenga1980shock}
Lyzenga, G.~A., and T.~J. Ahrens (1980), Shock temperature measurements in
  {Mg$_2$SiO$_4$} and {SiO$_2$} at high pressures, \textit{Geophysical Research
  Letters}, \textit{7}(2), 141--144.

\bibitem[{\textit{Matzen}(1997)}]{matzen1997z}
Matzen, M.~K. (1997), Z pinches as intense x-ray sources for high-energy
  density physics applications, \textit{Physics of Plasmas}, \textit{4}(5),
  1519--1527.

\bibitem[{\textit{McQueen et~al.}(1970)\textit{McQueen, Marsh, Taylor, Fritz,
  and Carter}}]{mcqueen1970equation}
McQueen, R., S.~Marsh, J.~Taylor, J.~Fritz, and W.~Carter (1970), The equation
  of state of solids from shock wave studies, \textit{High velocity impact
  phenomena}, \textit{293}, 294--417.

\bibitem[{\textit{Melosh}(2007)}]{melosh2007hydrocode}
Melosh, H. (2007), A hydrocode equation of state for {SiO$_2$},
  \textit{Meteoritics \& Planetary Science}, \textit{42}(12), 2079--2098.

\bibitem[{\textit{{Meng} et~al.}(2014)\textit{{Meng}, {Su}, {Rieke},
  {Stevenson}, {Plavchan}, {Rujopakarn}, {Lisse}, {Poshyachinda}, and
  {Reichart}}}]{Meng14}
{Meng}, H.~Y.~A., K.~Y.~L. {Su}, G.~H. {Rieke}, D.~J. {Stevenson},
  P.~{Plavchan}, W.~{Rujopakarn}, C.~M. {Lisse}, S.~{Poshyachinda}, and D.~E.
  {Reichart} (2014), {Large impacts around a solar-analog star in the era of
  terrestrial planet formation}, \textit{Science}, \textit{345}, 1032--1035,
  \doi{10.1126/science.1255153}.

\bibitem[{\textit{Miller et~al.}(2007)\textit{Miller, Boehly, Melchior,
  Meyerhofer, Celliers, Eggert, Hicks, Sorce, Oertel, and
  Emmel}}]{miller2007streaked}
Miller, J., T.~Boehly, A.~Melchior, D.~Meyerhofer, P.~Celliers, J.~Eggert,
  D.~Hicks, C.~Sorce, J.~Oertel, and P.~Emmel (2007), Streaked optical
  pyrometer system for laser-driven shock-wave experiments on {OMEGA},
  \textit{Review of Scientific Instruments}, \textit{78}(3), 034,903.

\bibitem[{\textit{Mosenfelder et~al.}(2007)\textit{Mosenfelder, Asimow, and
  Ahrens}}]{mosenfelder2007thermodynamic}
Mosenfelder, J.~L., P.~D. Asimow, and T.~J. Ahrens (2007), Thermodynamic
  properties of {Mg$_2$SiO$_4$} liquid at ultra-high pressures from shock
  measurements to 200 {GPa} on forsterite and wadsleyite, \textit{Journal of
  Geophysical Research: Solid Earth}, \textit{112}(B6).

\bibitem[{\textit{Mosenfelder et~al.}(2009)\textit{Mosenfelder, Asimow, Frost,
  Rubie, and Ahrens}}]{mosenfelder2009mgsio3}
Mosenfelder, J.~L., P.~D. Asimow, D.~J. Frost, D.~C. Rubie, and T.~J. Ahrens
  (2009), The {MgSiO$_3$} system at high pressure: {T}hermodynamic properties
  of perovskite, postperovskite, and melt from global inversion of shock and
  static compression data, \textit{Journal of Geophysical Research: Solid
  Earth}, \textit{114}(B1).

\bibitem[{\textit{Mysen and Kushiro}(1988)}]{mysen1988condensation}
Mysen, B.~O., and I.~Kushiro (1988), Condensation, evaporation, melting, and
  crystallization in the primitive solar nebula; experimental data in the
  system {MgO-SiO} 2-{H} 2 to 1.0 x10 $^{-9}$ bar and 1870 degrees {C} with
  variable oxygen fugacity, \textit{American Mineralogist}, \textit{73}(1-2),
  1--19.

\bibitem[{\textit{Nagahara et~al.}(1994)\textit{Nagahara, Kushiro, and
  Mysen}}]{nagahara1994evaporation}
Nagahara, H., I.~Kushiro, and B.~O. Mysen (1994), Evaporation of olivine: {L}ow
  pressure phase relations of the olivine system and its implication for the
  origin of chondritic components in the solar nebula, \textit{Geochimica et
  cosmochimica acta}, \textit{58}(8), 1951--1963.

\bibitem[{\textit{Nakajima and Stevenson}(2014)}]{nakajima2014investigation}
Nakajima, M., and D.~J. Stevenson (2014), Investigation of the initial state of
  the {M}oon-forming disk: {B}ridging {SPH} simulations and hydrostatic models,
  \textit{Icarus}, \textit{233}, 259--267.

\bibitem[{\textit{Pierazzo et~al.}(1997)\textit{Pierazzo, Vickery, and
  Melosh}}]{pierazzo1997reevaluation}
Pierazzo, E., A.~Vickery, and H.~Melosh (1997), A reevaluation of impact melt
  production, \textit{Icarus}, \textit{127}(2), 408--423.

\bibitem[{\textit{Presnall and Walter}(1993)}]{presnall1993melting}
Presnall, D.~C., and M.~J. Walter (1993), Melting of forsterite,
  {Mg$_2$SiO$_4$}, from 9.7 to 16.5 {GPa}, \textit{Journal of Geophysical
  Research: Solid Earth}, \textit{98}(B11), 19,777--19,783.

\bibitem[{\textit{Quintana et~al.}(2002)\textit{Quintana, Lissauer, Chambers,
  and Duncan}}]{quintana2002terrestrial}
Quintana, E.~V., J.~J. Lissauer, J.~E. Chambers, and M.~J. Duncan (2002),
  Terrestrial planet formation in the $\alpha$ {C}entauri system, \textit{The
  Astrophysical Journal}, \textit{576}(2), 982.

\bibitem[{\textit{Quintana et~al.}(2016)\textit{Quintana, Barclay, Borucki,
  Rowe, and Chambers}}]{quintana2016frequency}
Quintana, E.~V., T.~Barclay, W.~J. Borucki, J.~F. Rowe, and J.~E. Chambers
  (2016), The frequency of giant impacts on {E}arth-like worlds, \textit{The
  Astrophysical Journal}, \textit{821}(2), 126.

\bibitem[{\textit{Raymond et~al.}(2009)\textit{Raymond, O'Brien, Morbidelli,
  and Kaib}}]{raymond2009building}
Raymond, S.~N., D.~P. O'Brien, A.~Morbidelli, and N.~A. Kaib (2009), Building
  the terrestrial planets: Constrained accretion in the inner solar system,
  \textit{Icarus}, \textit{203}(2), 644--662.

\bibitem[{\textit{Rice et~al.}(1958)\textit{Rice, McQueen, and
  Walsh}}]{rice1958compression}
Rice, M., R.~G. McQueen, and J.~Walsh (1958), Compression of solids by strong
  shock waves, in \textit{Solid state physics}, vol.~6, pp. 1--63, Elsevier.

\bibitem[{\textit{Richet et~al.}(1993)\textit{Richet, Leclerc, and
  Benoist}}]{richet1993melting}
Richet, P., F.~Leclerc, and L.~Benoist (1993), Melting of forsterite and
  spinel, with implications for the glass transition of {Mg$_2$SiO$_4$} liquid,
  \textit{Geophysical Research Letters}, \textit{20}(16), 1675--1678.

\bibitem[{\textit{Robie and Hemingway}(1995)}]{robie1995thermodynamic}
Robie, R.~A., and B.~S. Hemingway (1995), \textit{Thermodynamic properties of
  minerals and related substances at 298.15 K and 1 bar (105 Pascals) pressure
  and at higher temperatures}, vol. 2131, US Government Printing Office.

\bibitem[{\textit{Robie et~al.}(1982)\textit{Robie, Hemingway, and
  Takei}}]{robie1982heat}
Robie, R.~A., B.~S. Hemingway, and H.~Takei (1982), Heat capacities and
  entropies of {Mg$_2$SiO$_4$}, {Mn$_2$SiO$_4$}, and {Co$_2$SiO$_4$} between 5
  and 380 {K}, \textit{American Mineralogist}, \textit{67}(5-6), 470--482.

\bibitem[{\textit{Root et~al.}(2015)\textit{Root, Mattsson, Cochrane, Lemke,
  and Knudson}}]{root2015shock}
Root, S., T.~R. Mattsson, K.~Cochrane, R.~W. Lemke, and M.~D. Knudson (2015),
  Shock compression response of poly (4-methyl-1-pentene) plastic to 985 {GPa},
  \textit{Journal of Applied Physics}, \textit{118}(20), 205,901.

\bibitem[{\textit{Root et~al.}(2018)\textit{Root, Townsend, Davies, Lemke,
  Bliss, Fratanduono, Kraus, Millot, Spaulding, Shulenburger, Stewart, and
  Jacobsen}}]{root2018forsterite}
Root, S., J.~P. Townsend, E.~J. Davies, R.~W. Lemke, D.~E. Bliss, D.~E.
  Fratanduono, R.~G. Kraus, M.~Millot, D.~K. Spaulding, L.~Shulenburger, S.~T.
  Stewart, and S.~B. Jacobsen (2018), The principal {H}ugoniot of forsterite to
  950 {GPa}, \textit{Geophysical Research Letters}.

\bibitem[{\textit{Root et~al.}(2019)\textit{Root, Townsend, and
  Knudson}}]{root2019FS}
Root, S., J.~P. Townsend, and M.~D. Knudson (2019), Shock compression of fused
  silica: An impedance matching standard, \textit{Journal of Applied Physics},
  \textit{126}(16), 165,901.

\bibitem[{\textit{Savage et~al.}(2007)\textit{Savage, Bennett, Bliss, Clark,
  Coats, Elizondo, LeChien, Harjes, Lehr, Maenchen
  et~al.}}]{savage2007overview}
Savage, M., L.~Bennett, D.~Bliss, W.~Clark, R.~Coats, J.~Elizondo, K.~LeChien,
  H.~Harjes, J.~Lehr, J.~Maenchen, et~al. (2007), An overview of pulse
  compression and power flow in the upgraded {Z} pulsed power driver, in
  \textit{Pulsed Power Conference, 2007 16th IEEE International}, vol.~2, pp.
  979--984, IEEE.

\bibitem[{\textit{Spielman et~al.}(1996)\textit{Spielman, Breeze, Deeney,
  Douglas, Long, Martin, Matzen, McDaniel, McGurn, Nash
  et~al.}}]{spielman1996pbfa}
Spielman, R., S.~Breeze, C.~Deeney, M.~Douglas, F.~Long, T.~Martin, M.~Matzen,
  D.~McDaniel, J.~McGurn, T.~Nash, et~al. (1996), {PBFA Z}: {A} 20-{MA} z-pinch
  driver for plasma radiation sources, in \textit{High-Power Particle Beams,
  1996 11th International Conference on}, vol.~1, pp. 150--153, IET.

\bibitem[{\textit{Stewart et~al.}(2019{\natexlab{a}})\textit{Stewart, Carter,
  Davies, Lock, Kraus, Root, Petaev, and Jacobsen}}]{stewart2019impact}
Stewart, S., P.~Carter, E.~Davies, S.~Lock, R.~Kraus, S.~Root, M.~Petaev, and
  S.~Jacobsen (2019{\natexlab{a}}), Impact vapor plume expansion and
  hydrodynamic collapse in the solar nebula., in \textit{Lunar and Planetary
  Science Conference}, vol.~50.

\bibitem[{\textit{Stewart and Ahrens}(2005)}]{stewart2005shock}
Stewart, S.~T., and T.~J. Ahrens (2005), Shock properties of {H$_2$O} ice,
  \textit{Journal of Geophysical Research: Planets}, \textit{110}(E3).

\bibitem[{\textit{Stewart and Leinhardt}(2012)}]{stewart2012collisions}
Stewart, S.~T., and Z.~M. Leinhardt (2012), Collisions between
  gravity-dominated bodies. {II}. {T}he diversity of impact outcomes during the
  end stage of planet formation, \textit{The Astrophysical Journal},
  \textit{751}(1), 32.

\bibitem[{\textit{Stewart et~al.}(2019{\natexlab{b}})\textit{Stewart, Davies,
  Duncan, Lock, Root, Townsend, Kraus, Caracas, and
  Jacobsen}}]{stewart2019improvements}
Stewart, S.~T., E.~J. Davies, M.~S. Duncan, S.~J. Lock, S.~Root, J.~P.
  Townsend, R.~G. Kraus, R.~Caracas, and S.~B. Jacobsen (2019{\natexlab{b}}),
  The shock physics of giant impacts: {K}ey requirements for the equations of
  state, \textit{arXiv preprint arXiv:1910.04687v2}.

\bibitem[{\textit{{Su} et~al.}(2019)\textit{{Su}, {Jackson}, {G{\'a}sp{\'a}r},
  {Rieke}, {Dong}, {Olofsson}, {Kennedy}, {Leinhardt}, {Malhotra}, {Hammer},
  {Meng}, {Rujopakarn}, {Rodriguez}, {Pepper}, {Reichart}, {James}, and
  {Stassun}}}]{Su19}
{Su}, K. Y.~L., A.~P. {Jackson}, A.~{G{\'a}sp{\'a}r}, G.~H. {Rieke}, R.~{Dong},
  J.~{Olofsson}, G.~M. {Kennedy}, Z.~M. {Leinhardt}, R.~{Malhotra},
  M.~{Hammer}, H.~Y.~A. {Meng}, W.~{Rujopakarn}, J.~E. {Rodriguez},
  J.~{Pepper}, D.~E. {Reichart}, D.~{James}, and K.~G. {Stassun} (2019),
  {Extreme Debris Disk Variability: Exploring the Diverse Outcomes of Large
  Asteroid Impacts During the Era of Terrestrial Planet Formation}, \textit{The
  Astronomical Journal}, \textit{157}(5), 202, \doi{10.3847/1538-3881/ab1260}.

\bibitem[{\textit{Thomas and Asimow}(2013)}]{thomas2013direct}
Thomas, C.~W., and P.~D. Asimow (2013), Direct shock compression experiments on
  premolten forsterite and progress toward a consistent high-pressure equation
  of state for {CaO-MgO-Al$_2$O$_3$-SiO$_2$-FeO} liquids, \textit{Journal of
  Geophysical Research: Solid Earth}, \textit{118}(11), 5738--5752.

\bibitem[{\textit{Thompson}(1990)}]{thompson1990aneos}
Thompson, S.~L. (1990), {ANEOS} analytic equations of state for shock physics
  codes input manual, \textit{Tech. rep.}, Sandia National Labs., Albuquerque,
  NM (USA).

\bibitem[{\textit{Thompson and Lauson}(1974)}]{thompson1974improvements}
Thompson, S.~L., and H.~S. Lauson (1974), {Improvements in the Chart D
  radiation-hydrodynamic CODE III: Revised analytic equations of state},
  \textit{Tech. Rep. SC-RR--71-0714}, Sandia Labs., Albuquerque, NM.

\bibitem[{\textit{Tonks and Melosh}(1993)}]{tonks1993magma}
Tonks, W.~B., and H.~J. Melosh (1993), Magma ocean formation due to giant
  impacts, \textit{Journal of Geophysical Research: Planets}, \textit{98}(E3),
  5319--5333.

\bibitem[{\textit{Walsh et~al.}(2011)\textit{Walsh, Morbidelli, Raymond,
  O'brien, and Mandell}}]{walsh2011low}
Walsh, K.~J., A.~Morbidelli, S.~N. Raymond, D.~P. O'brien, and A.~M. Mandell
  (2011), A low mass for {M}ars from {J}upiter$\lq$s early
  gas-driven migration, \textit{Nature}, \textit{475}(7355), 206.

\bibitem[{\textit{Watt and Ahrens}(1983)}]{watt1983shock}
Watt, J.~P., and T.~J. Ahrens (1983), Shock compression of single-crystal
  forsterite, \textit{Journal of Geophysical Research: Solid Earth},
  \textit{88}(B11), 9500--9512.

\bibitem[{\textit{Wood}(2000)}]{wood2000pressure}
Wood, J.~A. (2000), Pressure and temperature profiles in the solar nebula, in
  \textit{From Dust to Terrestrial Planets}, pp. 87--93, Springer.

\end{thebibliography}

\clearpage
\begin{thebibliography}{8}
\providecommand{\natexlab}[1]{#1}
\expandafter\ifx\csname urlstyle\endcsname\relax
  \providecommand{\doi}[1]{doi:\discretionary{}{}{}#1}\else
  \providecommand{\doi}{doi:\discretionary{}{}{}\begingroup
  \urlstyle{rm}\Url}\fi

\bibitem[{\textit{Carter et~al.}(2015)\textit{Carter, Leinhardt, Elliott,
  Walter, and Stewart}}]{carter2015compositional}
Carter, P.~J., Z.~M. Leinhardt, T.~Elliott, M.~J. Walter, and S.~T. Stewart
  (2015), Compositional evolution during rocky protoplanet accretion,
  \textit{The Astrophysical Journal}, \textit{813}(1), 72.

\bibitem[{\textit{de~Koker et~al.}(2008)\textit{de~Koker, Stixrude, and
  Karki}}]{de2008thermodynamics}
de~Koker, N.~P., L.~Stixrude, and B.~B. Karki (2008), Thermodynamics,
  structure, dynamics, and freezing of {Mg$_2$SiO$_4$} liquid at high pressure,
  \textit{Geochimica et Cosmochimica Acta}, \textit{72}(5), 1427--1441.

\bibitem[{\textit{Knudson and Desjarlais}(2013)}]{knudson2013adiabatic}
Knudson, M., and M.~Desjarlais (2013), Adiabatic release measurements in
  $\alpha$-quartz between 300 and 1200 {GPa}: {C}haracterization of
  $\alpha$-quartz as a shock standard in the multimegabar regime,
  \textit{Physical Review B}, \textit{88}(18), 184,107.

\bibitem[{\textit{Kraus et~al.}(2012)\textit{Kraus, Stewart, Swift, Bolme,
  Smith, Hamel, Hammel, Spaulding, Hicks, Eggert et~al.}}]{kraus2012shock}
Kraus, R., S.~Stewart, D.~Swift, C.~Bolme, R.~Smith, S.~Hamel, B.~Hammel,
  D.~Spaulding, D.~Hicks, J.~Eggert, et~al. (2012), Shock vaporization of
  silica and the thermodynamics of planetary impact events, \textit{Journal of
  Geophysical Research: Planets}, \textit{117}(E9).

\bibitem[{\textit{Mysen and Kushiro}(1988)}]{mysen1988condensation}
Mysen, B.~O., and I.~Kushiro (1988), Condensation, evaporation, melting, and
  crystallization in the primitive solar nebula; experimental data in the
  system {MgO-SiO} 2-{H} 2 to 1.0 x10 $^{-9}$ bar and 1870 degrees {C} with
  variable oxygen fugacity, \textit{American Mineralogist}, \textit{73}(1-2),
  1--19.

\bibitem[{\textit{Richet et~al.}(1982)\textit{Richet, Bottinga, Denielou,
  Petitet, and Tequi}}]{richet1982thermodynamic}
Richet, P., Y.~Bottinga, L.~Denielou, J.~Petitet, and C.~Tequi (1982),
  Thermodynamic properties of quartz, cristobalite and amorphous {SiO$_2$}:
  {D}rop calorimetry measurements between 1000 and 1800 {K} and a review from 0
  to 2000 {K}, \textit{Geochimica et Cosmochimica Acta}, \textit{46}(12),
  2639--2658.

\bibitem[{\textit{Root et~al.}(2018)\textit{Root, Townsend, Davies, Lemke,
  Bliss, Fratanduono, Kraus, Millot, Spaulding, Shulenburger, Stewart, and
  Jacobsen}}]{root2018forsterite}
Root, S., J.~P. Townsend, E.~J. Davies, R.~W. Lemke, D.~E. Bliss, D.~E.
  Fratanduono, R.~G. Kraus, M.~Millot, D.~K. Spaulding, L.~Shulenburger, S.~T.
  Stewart, and S.~B. Jacobsen (2018), The principal {H}ugoniot of forsterite to
  950 {GPa}, \textit{Geophysical Research Letters}.

\bibitem[{\textit{Thomas and Asimow}(2013)}]{thomas2013direct}
Thomas, C.~W., and P.~D. Asimow (2013), Direct shock compression experiments on
  premolten forsterite and progress toward a consistent high-pressure equation
  of state for {CaO-MgO-Al$_2$O$_3$-SiO$_2$-FeO} liquids, \textit{Journal of
  Geophysical Research: Solid Earth}, \textit{118}(11), 5738--5752.

\end{thebibliography}
\end{document}